\documentclass[10pt]{article}
\usepackage[T1]{fontenc}
\usepackage{amsmath,amsfonts,amsthm,mathrsfs,amssymb}
\usepackage{cite}
\usepackage{multirow}
\usepackage{graphicx,float}
\usepackage{subfigure}
\usepackage{placeins}
\usepackage{color}
\usepackage{indentfirst}
\usepackage[bookmarks=true]{hyperref}
\numberwithin{equation}{section}
\newcommand{\R}{{\mathbb R}}

\newcommand{\be}{\begin{eqnarray}}
\newcommand{\ben}{\begin{eqnarray*}}
\newcommand{\en}{\end{eqnarray}}
\newcommand{\enn}{\end{eqnarray*}}

\newcommand{\pa}{\partial}

\newcommand{\curl}{{\rm curl\,}}

\newcommand{\divv}{{\rm div\,}}

\newcommand{\G}{\Gamma}

\newtheorem{theorem}{Theorem}[section]

\newtheorem{remark}[theorem]{Remark}

\definecolor{rot}{rgb}{1.000,0.000,0.000}
\definecolor{rot1}{rgb}{0.000,0.000,0.000}
\begin{document}
\renewcommand{\theequation}{\arabic{section}.\arabic{equation}}
\begin{titlepage}
  \title{A Windowed Green Function method for  \\
    elastic scattering problems on a half-space}

\author{Oscar P. Bruno\thanks{Department of Computing \& Mathematical Sciences, California Institute of Technology, 1200 East California Blvd., CA 91125, United States. Email:{\tt obruno@caltech.edu}}\;\;and
Tao Yin\thanks{Department of Computing \& Mathematical Sciences, California Institute of Technology, 1200 East California Blvd., CA 91125, United States. Email:{\tt taoyin89@caltech.edu}}}
\end{titlepage}
\maketitle
%\vspace{.2in}

\begin{abstract}
  This paper presents a windowed Green function (WGF) method for the
  numerical solution of problems of elastic scattering by
  ``locally-rough surfaces'' (i.e., local perturbations of a half
  space), under either Dirichlet or Neumann boundary conditions, and
  in both two and three spatial dimensions. The proposed WGF method
  relies on an integral-equation formulation based on the {\em
    free-space Green function}, together with smooth operator
  windowing (based on a ``slow-rise'' windowing function) and
  efficient high-order singular-integration methods. The approach
  avoids the evaluation of the expensive layer Green function for
  elastic problems on a half-space, and it yields uniformly fast
  convergence for all incident angles. Numerical experiments for both
  two and three dimensional problems are presented, demonstrating the
  accuracy and super-algebraically fast convergence of the proposed
  method as the window-size grows.  {\bf Keywords:} Elastic wave,
  half-space, windowed Green function, boundary integral equation
\end{abstract}

\section{Introduction}
\label{sec:1}

In view of their great importance in diverse areas of applications,
the problems of scattering by unbounded rough-surfaces, including
scattering of acoustic, electromagnetic, and elastic waves, have
attracted the interest of physicists, engineers, and mathematicians
for many years. Specifically, simulations concerning elastic
half-space problems (in which material interfaces are everywhere
planar, except for bounded regions which may contain arbitrarily
complex structures) play essential roles in the investigation of
earthquakes, non-destructive testing of materials, and energy
production from natural gas and geothermal
sources~\cite{A73,AR02,S09}. This paper introduces an efficient
high-order integral solver for problems of this type. More precisely,
this paper presents an efficient and accurate methodology, based on
surface integral equations over the material interfaces, for the
problem of elastic wave scattering over a
half-space~\cite{A00,A01,A02,EH12,EH15}. In particular, the method is
applicable to configurations in which the scattering boundary is a
combination of an unbounded flat surface and local (bounded)
non-planar surface perturbations and/or bounded elastic scatterers. Unlike
the volumetric discretization methods for these problems, the
boundary integral equation (BIE) approach~\cite{HW08,N01} only
requires discretization of regions of lower dimensionality, and it
automatically enforces the radiation condition at infinity. In
conjunction with adequate acceleration techniques (see
e.g.~\cite{BK01,CBS08,L09}) for the associated matrix-vector products
and Krylov-subspace linear algebra solver such as GMRES, the BIE
method can provide fast, high-order solvers even for problems of high
frequency.

Two main integral equation approaches have been used for scattering
problems on a half-space. One is based on the layer Green function
(LGF)~\cite{A00,CB14,DGN10,DMN11}---which automatically enforces the
relevant boundary conditions on the unbounded flat surfaces and thus
reduces the scattering problems to integral equations on the
defects. It should be pointed out that the Dirichlet and Neumann
cases, for which the layer Green function is trivially calculated in
the acoustic case, require Fourier-transform based layer Green
function in the elastic case. A second approach relies on integral
equations imposed on the complete unbounded
surface~\cite{BLPT16,DM97,CGK02,P16}. The potential benefits of the
second approach arise from its use of the free-space Green-function
kernel, whose evaluation cost is much lower, by orders of magnitude,
than the LGF evaluation cost---since evaluation of a single value of
the LGF requires computation of challenging Fourier integrals
containing highly-oscillatory integrands over infinite integration
intervals; see e.g.~\cite[Eq. (2.27)]{A00} and~\cite[Eq. (26)]{CB14}.

The integral equations based on the free-space Green function, on the
other hand, are posed on the complete unbounded interface, and they
therefore require, for computational purposes, use of a
domain-truncation strategy of some sort---which raises questions with
regard to selection of suitable truncation radii and the potentially
large number of required unknowns~\cite{AA04,CB13}. For the elastic
scattering problems in a half-space, a direct-truncation approach is
discussed in~\cite{CBS08,CB13,GCBS12} (in which, significantly, only
normal-incidence problems are considered). The examples considered in
these papers suggest that a truncation radius equal to three to five
times the radius of the surface irregularity yields acceptable
accuracy for normal-incidence problems.  However, as illustrated in
Section~\ref{sec:3.1} for a related approach, this truncation strategy
requires, for a given accuracy, use of larger and larger truncated
domains as the incidence angles depart from normal, with required
inclusion of planar sections that grow beyond all finite bounds as the
incidence angle approaches grazing.

The present paper proposes a novel truncation approach, called the
windowed Green function (WGF) method, for the problem of elastic
scattering on a half-space. The WGF method has previously been found
effective in the contexts of acoustic and electromagnetic scattering
by periodic structures~\cite{BD14,BSTV16,M07}, multiply-layered
media~\cite{BLPT16,BP17,P16}, waveguide structures~\cite{BGP17} and
long-range volumetric propagation~\cite{CBA09}. On the basis of
certain ``slow-rise'' windowing functions $\widetilde{w}_A$, the WGF
method we propose here truncates the original integral equations over
unbounded surfaces to integration domains that include the surface
defects and appropriate portions of the flat interfaces. As for the
direct-truncation method, however, straightforward windowing of the
scattering integral operators requires use of windowed regions that
grow without bound, to meet a fixed error tolerance, as the incidence
angle approaches grazing (see Section~\ref{sec:3.1}). To overcome this
difficulty, the proposed method introduces a correction that smoothly
merges the unknown density values in the original integral equations
with values of the corresponding solutions of scattering by a
perfectly flat surface. This modification allows the WGF method to
yield super-algebraically accurate approximations of the exact
infinite-domain solutions throughout the region wherein the window
function equals one. As demonstrated via a variety of numerical
examples in Sections~\ref{sec:3.2} and \ref{sec:5}, the corrected WGF
method provides uniformly fast convergence, over all incident angles,
as the support of windowing function grows.

It is relevant to recall that the classical integral operators of
elasticity theory, which are presented in Section~\ref{sec:2.2}, are
strongly singular operators defined in terms of Cauchy principal-value
integrals. But the strong singularity of these operators stems from
differentiation of certain weakly singular kernels and thus, as shown
in~\cite{BXY19,YHX17} using an integration-by-parts argument, the
operators can be re-expressed as compositions of weakly-singular
integral operators (with kernels expressed in terms of the free-space
elastic Green function $E$ and its normal derivatives, at least for
smooth boundaries), as well as certain tangential ``G\"unter
derivatives'' weakly-singular free-space elastic Green function (which
result in strongly-singular kernels). In detail, focusing on problems
of scattering by bounded obstacles, those references utilize an
integration-by-parts procedure to recast the action of a
strongly-singular operator on a given density in terms of the action
of an associated weakly-singular operator applied to certain
derivatives of the density. In the present context, the strongly
singular operators (equations~\eqref{KWA} and \eqref{KPWA}) are posed
on a surface with boundary (the boundary of the computational
integration domain), but, as indicated in Section~\ref{sec:4.1}, no
boundary contributions result in the integration-by-parts process in
this case either, since the window function we use, which is part of
the operator integrand, vanishes at the boundary of the integration
domain.

The overall proposed procedure thus reduces the operator evaluation
problem to evaluation of weakly singular operators and tangential
differentiation of surface densities. The weakly-singular integration
problem is tackled in this paper by means of the Chebyshev-based
rectangular-polar discretization methodology introduced
recently~\cite{BG18,BY19}---which can be readily applied in
conjunction with geometry descriptions given by a set of
non-overlapping logically-quadrilateral patches, and which, therefore,
makes the algorithm particularly well suited for treatment of complex
geometries. The needed tangential differentiations, in turn, can
easily be produced by means of differentiation of corresponding
truncated Chebyshev expansions, with evaluation either via FFT or, for
sufficiently small expansions, via direct summation.

This paper is organized as follows. Section~\ref{sec:2} describes the
half-space elastic scattering problems under consideration, and it
presents corresponding BIEs based on the free-space Green
function. Section~\ref{sec:3} then presents the proposed 3D WGF
methodology, including a description of the windowed integral
operators and a preliminary windowed integral formulation
(Section~\ref{sec:3.1}), as well as a ``corrected'' windowed integral
formulation which is uniformly accurate for all incident angles, up to
grazing (Section~\ref{sec:3.2}). Section~\ref{sec:4} introduces the
proposed high order operator discretization methods we use in our 2D
implementation; the 3D operator discretization methods we use are
described in~\cite{BG18,BY19}. A variety of numerical examples in 2D
and 3D, finally, are presented in Section~\ref{sec:5}---demonstrating
the accuracy and efficiency of the overall proposed approach.

\section{Preliminaries}
\label{sec:2}
\subsection{Elastic scattering problems}
\label{sec:2.1}

\begin{figure}[ht]
\centering
\includegraphics[scale=0.4]{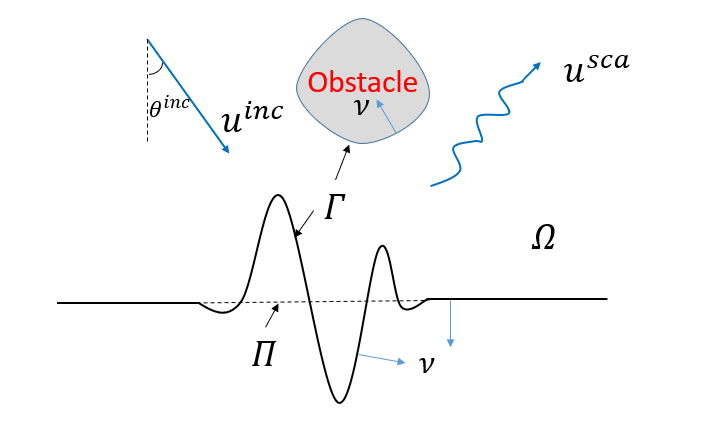}
\caption{Problem of scattering by a locally perturbed elastic
  half-space $\Omega\subseteq \mathbb{R}^d$ ($d=2$ or $d=3$).}
\label{model}
\end{figure}

Let $\Omega\in\R^d,d=2,3$ denote an unbounded connected open set as
illustrated in Figure~\ref{model} which, in particular, satisfies \ben
U_{f_+}\subset\Omega\subset U_{f_-},\quad
U_{f_\pm}:=\{x=(x_1,\dots,x_d)\in\R^d: x_d>f_\pm\} \enn for certain
constants $f_-<f_+$.  Let $\Gamma:=\partial\Omega$ denote the
unbounded rough surface which, in addition to the unbounded flat
surface, encompasses either a local defect on the flat surface
$\Pi:=\{x\in\R^d: x_d=0\}$ or a bounded obstacle in $U_0$, or a
combination thereof. Assume that the unbounded domain $\Omega$ is
occupied by a linear isotropic and homogeneous elastic medium
characterized by the Lam\'e constants $\lambda,\mu$ ($\mu>0$,
$d\lambda+2\mu>0$) and the mass density $\rho>0$. Denote by $\omega$
the frequency and by \ben k_s := \omega\sqrt{\rho/\mu},\quad k_p =
\omega \sqrt{\rho/(\lambda + 2\mu)} \enn the shear and compressional
wave numbers, respectively. For definiteness, throughout this paper
consider cases in which the incident field $u^{inc}$ equals a plane
pressure wave, but other types of boundary conditions, including plane
share waves, can be treated similarly. A plane pressure wave is given
by the expression \be
\label{incp}
u^{inc}=d^{inc}e^{ik_px\cdot d^{inc}},
\en
where
\ben
d^{inc}=\begin{pmatrix}
\sin\theta^{inc}\\ -\cos\theta^{inc}
\end{pmatrix}\quad\mbox{in}\quad\mbox{2D}\quad\mbox{and}\quad
d^{inc}=\begin{pmatrix}
\sin\theta^{inc}\\ 0\\ -\cos\theta^{inc}
\end{pmatrix}\quad\mbox{in}\quad\mbox{3D}
\enn
represents the incident versor direction and $\theta^{inc}$ denotes the incident angle satisfying $|\theta^{inc}|<\pi/2$. Suppressing the time-harmonic dependence $e^{-i\omega t}$, the scattered displacement field $u^\mathrm{scat}$ can be modeled by time-harmonic Navier equation
\be
\label{navier}
\Delta^*u^\mathrm{scat}+\rho\omega^2u^\mathrm{scat}=0 \quad\mbox{in}\quad\Omega, \en
with either Dirichlet boundary conditions \ben u^\mathrm{scat}=
-u^{inc}\quad\mbox{on}\quad\Gamma, \enn or Neumann boundary conditions
\ben T(\pa,\nu)u^\mathrm{scat}=-T(\pa,\nu)u^{inc} \quad\mbox{on}\quad\Gamma.
\enn and with an upward propagating radiation condition at infinity
(UPRC)~\cite{AH05,EH12,CGK02}. Here $\Delta^{*}$ and $T(\pa,\nu)$
denote the Lam\'e operator \ben
\label{LameOper}
\Delta^* := \mu\,\mbox{div}\,\mbox{grad} + (\lambda +
\mu)\,\mbox{grad}\, \mbox{div}\,, \enn and the traction operator \be
\label{stress}
T(\pa,\nu)u:=2 \mu \, \partial_{\nu} u + \lambda \, \nu \, \divv u+\mu
\nu\times \curl u, \en respectively, where $\nu$ and
$\partial_\nu:=\nu\cdot\nabla$ denote the outward unit normal to $\G$
and the normal derivative, respectively.

\begin{remark}
  \label{remark_exact}
  In the case $\Gamma=\Pi$, for which no local defect or obstacles
  exist, the exact solution in 2D (with a similar result in 3D) under
  an incident plane wave~\eqref{incp} is given by
  \ben u^\mathrm{scat}_f=A_p\begin{bmatrix} \sin\theta^{inc}\\
    \cos\theta^{inc}
\end{bmatrix}e^{ik_p(x_1\sin\theta^{inc}+x_2\cos\theta^{inc})}+ A_s\begin{bmatrix}
-\cos\theta_s\\ \sin\theta_s
\end{bmatrix}e^{ik_p(x_1\sin\theta_s+x_2\cos\theta_s)} \enn where
$k_s\sin\theta_s=k_p\sin\theta^{inc}, |\theta_s|<\pi/2$. The boundary
conditions on $\Pi$ tell us that the factors $A_p$ and $A_s$ can be
obtained from the linear systems \ben
\begin{bmatrix}
\sin\theta^{inc} & -\cos\theta_s\\ \cos\theta^{inc} & \sin\theta_s
\end{bmatrix}\begin{bmatrix}
A_p\\ A_s
\end{bmatrix}=\begin{bmatrix}
-\sin\theta^{inc} \\ \cos\theta^{inc}
\end{bmatrix},
\enn
and
\ben
\begin{bmatrix}
-2i\mu k_p\sin\theta^{inc}\cos\theta^{inc} & 2i\mu k_s\cos^2\theta_s-i\mu k_s\\
-2i\mu k_p\cos^2\theta^{inc}-i\lambda k_p & -2i\mu k_s\sin\theta_s\cos\theta_s
\end{bmatrix}\begin{bmatrix}
A_p\\ A_s
\end{bmatrix}=\begin{bmatrix}
-2i\mu k_p\sin\theta^{inc}\cos\theta^{inc} \\ 2i\mu k_p\cos^2\theta^{inc}+i\lambda k_p
\end{bmatrix}
\enn
for the Dirichlet and Neumann problems, respectively.
\end{remark}

\subsection{Boundary integral equation based on the free-space Green function}
\label{sec:2.2}

As is known~\cite{CGK02}, the scattered field $u^\mathrm{scat}$ admits the representation
\be
\label{scawaverepre}
u^\mathrm{scat}(x)= \mathcal{S}[Tu^\mathrm{scat}](x)-\mathcal{D}[u^\mathrm{scat}](x),\quad x\in\Omega,
\en
where, letting
\ben
G_{k}(x,y)=\begin{cases}
\frac{i}{4}H_0^{(1)}(k|x-y|), & d=2,\cr
\frac{e^{ik|x-y|}}{4\pi|x-y|}, & d=3,
\end{cases}, \enn and \ben
E(x,y)=\frac{1}{\mu}G_{k_s}(x,y)I+\frac{1}{\rho\omega^2}
\nabla_x\nabla_x^\top \left[G_{k_s}(x,y)-G_{k_p}(x,y)\right], \enn
denote the free-space Green functions for the Helmholtz equation (with
wavenumber $k$) and the Navier equation (with wavenumbers $k_s$ and
$k_p$), respectively, $\mathcal{S}$ and $\mathcal{D}$ denote the
single- and double-layer potentials \be
\label{singlelayer}
\mathcal{S}[\phi](x)&=&\int_\Gamma E(x,y)\phi(y)ds_y,\quad x\in\Omega,\\
\label{doublelayer}
\mathcal{D}[\phi](x)&=&\int_\Gamma
(T(\pa_y,\nu_y)E(x,y))^\top\phi(y)ds_y,\quad x\in\Omega. \en For
$|\theta^{inc}|<\pi/2$ the incident field $u^{inc}$ satisfies \be
\label{incwaverepre}
0=\mathcal{S}[Tu^{inc}](x)-\mathcal{D}[u^{inc}](x),\quad x\in\Omega.
\en
Taking the limit as $x\rightarrow\Gamma$ using well-known jump relations~\cite{HW08}, and applying the boundary conditions, we  obtain the BIE
\be
\label{DBIE}
-\frac{1}{2}\phi+K'[\phi]=-Tu^{inc} \quad\mbox{on}\quad\Gamma, \quad \phi=Tu^{tot},
\en
for the Dirichlet problem, and the BIE
\be
\label{NBIE}
\frac{1}{2}\psi+K[\psi]=u^{inc} \quad\mbox{on}\quad\Gamma, \quad
\psi=u^{tot}, \en for the Neumann problem, where
$u^{tot}:=u^\mathrm{scat}+u^{inc}$ denotes the total field, and where \be
\label{dlo}
K[\phi](x)&=&\int_\Gamma (T(\pa_y,\nu_y)E(x,y))^\top\phi(y)ds_y,\quad x\in\Gamma,\\
\label{tdlo}
K'[\psi](x)&=&\int_\Gamma T(\pa_x,\nu_x)E(x,y)\psi(y)ds_y,\quad
x\in\Gamma, \en denote the double-layer and transpose double-layer
integral operators (which are only defined in the sense of Cauchy
principle value).

%\begin{remark}
\subsection{Boundary integral equation based on the layer Green function}
\label{sec:2.3}

In addition to the ``free-space Green function'' BIEs presented in the
previous section we also mention, for reference, the corresponding
``bounded-surface'' BIEs based on the layer Green function.  The LGF
$\widetilde{E}(\cdot,y)$ satisfies \ben \Delta^*
\widetilde{E}(\cdot,y)+\rho\omega^2\widetilde{E}(\cdot,y)
=-\delta_y(\cdot) \quad\mbox{in}\quad U_0, \enn as well as homogeneous
Dirichlet or Neumann boundary condition on the flat surface $\Pi$ and
the UPRC at infinity. As is known~\cite{A00,CB14,DGN10,DMN11},
$\widetilde{E}$ can be expressed explicitly in terms of Fourier
integrals.

Integral equations posed on bounded
surfaces can be obtained, on the basis of the LGF, for the problem of
scattering by the unbounded surface $\Gamma$. Indeed, letting
$u^s:=u^\mathrm{scat}-u_f^\mathrm{scat}$, it follows that $u^s=0$ and $Tu^s=0$ on
$\Gamma\cap\Pi$ for Dirichlet and Neumann problems, respectively. In
view of the homogeneous boundary condition satisfied by the LGF on
$\Pi$, it follows from Green's formula that the solution $u^s$ can be
expressed in the form \be
\label{LGMscawaverepre}
u^s(x)=\widetilde{\mathcal{S}}[Tu^s](x)-\widetilde{\mathcal{D}}[u^s](x),\quad
x\in\Omega, \en where the single-layer potential
$\widetilde{\mathcal{S}}$ and double-layer potential
$\widetilde{\mathcal{D}}$ are given by \ben
\label{LGMsinglelayer}
\widetilde{\mathcal{S}}[\phi](x)&=&\int_{\Gamma\backslash\Pi} \widetilde{E}(x,y)\phi(y)ds_y,\quad x\in\Omega,\\
\label{LGMdoublelayer}
\widetilde{\mathcal{D}}[\phi](x)&=&\int_{\Gamma\backslash\Pi}
(T(\pa_y,\nu_y)\widetilde{E}(x,y))^\top\phi(y)ds_y,\quad x\in\Omega.
\enn Letting $x$ approach to the inhomogeneity $\Gamma\backslash\Pi$,
the BIEs on $\Gamma\backslash\Pi$ \ben
\widetilde{S}[Tu^s]&=& -\frac{1}{2}u^s+ \widetilde{K}[u^s],\\
\frac{1}{2}Tu^s+ \widetilde{K'}[Tu^s]&=& \widetilde{N}[u^s], \enn
result, where, for $x\in\Gamma\backslash\Pi$ we have set
\ben
\widetilde{S}[\phi](x)=\int_{\Gamma\backslash\Pi} \widetilde{E}(x,y)\phi(y)ds_y,\quad \widetilde{K}[\phi](x)=\int_{\Gamma\backslash\Pi} (T(\pa_y,\nu_y)\widetilde{E}(x,y))^\top\phi(y)ds_y,\\
\widetilde{N}[\phi](x)=\int_{\Gamma\backslash\Pi}
T(\pa_x,\nu_x)(T(\pa_y,\nu_y)\widetilde{E}(x,y))^\top\phi(y)ds_y,\quad
\widetilde{K'}[\phi](x)=\int_{\Gamma\backslash\Pi}
T(\pa_x,\nu_x)\widetilde{E}(x,y)\psi(y)ds_y.  \enn

It is important to note that the boundary integrals operators arising
from use of the LGF are posed on the bounded surface
$\Gamma\backslash\Pi$ (the local defect), and, in particular, their
numerical implementation does not require truncation of an infinite
physical domain. However, the evaluation of the elastic layer Green
function is much more expensive than the evaluation of the elastic
free space Green function~\cite{CB13,DGN10}---which motivated our
search for accurate and efficient truncation strategies.
%\end{remark}

\section{Windowed Green function method (WGF)}
\label{sec:3}

This section proposes the  WGF method for truncation of the
integral equations~(\ref{DBIE}) and (\ref{NBIE}). The WGF
method ensures superalgebraically fast convergence as the window size
is increased, and uniform accuracy at fixed computational cost for
arbitrary angles of incidence.
\vspace{0.5cm}

\subsection{Slow-rise windowing function and preliminary considerations}
\label{sec:3.1}

In order to achieve effective domain truncation, a smooth ``slow-rise''
windowing function
\ben
w_A(t)=\eta(t/A;c,1), \enn where \ben \eta(t;t_0,t_1)=\begin{cases} 1,
  & |t|\le t_0, \cr e^{\frac{2e^{-1/u}}{u-1}}, & t_0<|t|<t_1,
  u=\frac{|t|-t_0}{t_1-t_0}, \cr 0, & |t|\ge t_1.
\end{cases}
\enn was introduced in \cite{BLPT16,P16} in the context of acoustic
layered-media scattering. The function vanishes outside an interval of
length $2A$, it equals one in a region around the origin which grows
linearly with $A$, and it has a slow rise: all of its derivatives
tends to zero uniformly as $A\to\infty$. The width $2A>0$ of the
support of the windowed function $w_A$ should be selected so as to
ensure that $1-w_A(x_1)$ vanishes on the local defect
$\Gamma\backslash\Pi$, and should be additionally be large enough to
meet a given error tolerance.

Utilizing the windowing function
\ben
\widetilde{w}_A(x)=\begin{cases}
w_A(x_1) , & \mbox{in}\quad 2D,\cr
w_A(x_1)w_A(x_2), & \mbox{in}\quad 3D,
\end{cases}
\enn
we obtain the preliminary windowed version
\be
\label{WDBIE1}
-\frac{1}{2}\phi^*+K'[\widetilde{w}_A\phi^*]=-Tu^{inc}
\quad\mbox{on}\quad\Gamma_A \en of equation (\ref{DBIE}), where
$\Gamma_A$ denotes the part of the surface $\Gamma$ that
$\widetilde{w}_A(x)\ne 0$. Unfortunately, however, this formulation is
not uniformly accurate with respect to the angle of incidence.

\begin{figure}[ht]
\centering
\includegraphics[scale=0.3]{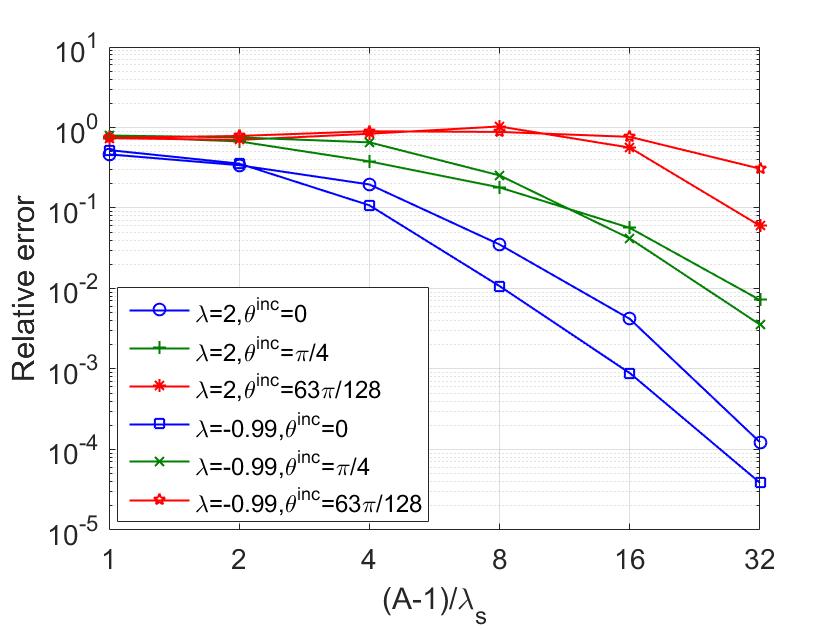}
\caption{Relative errors $\epsilon_\infty$ (equation~\eqref{rel_error}
  below) in the total field resulting from the preliminary WGF method
  for the Dirichlet problem of scattering by a semi-circular
  bump. Clearly, the preliminary WGF approach is not uniformly
  accurate as grazing incidence is approached.}
\label{Table3.1.1}
\end{figure}

To demonstrate the difficulty we consider the Dirichlet problem of
scattering of an incident wave $u^{inc}$ by a semi-circular bump of
radius $r=1$ in 2D with $\mu=1$, $\rho=1$ and $\omega=20$. The
integral equation (\ref{WDBIE1}) was discretized on the basis of the
high-order discretization approach introduced in Section \ref{sec:4}
with refinement exponent $p=4$
(Section~\ref{sec:4.2}). Figure~\ref{Table3.1.1} displays the relative
errors obtained in the total field \ben
u^{tot}(x)=u^{inc}(x)+\mathcal{S}[W_A\phi^*](x), \enn on the line
segment $\{x\in\R^2: -1\le x_1\le 1, x_2=2\}$ for two values of
$\lambda$ and and under various incidence angles. The errors displayed
in Figure~\ref{Table3.1.1} were evaluated by comparison with a
highly-resolved numerical solution for a large value of $A$.  The
results show that the direct windowing approach embodied in
(\ref{WDBIE1}) requires, for a given accuracy, increasingly large
truncated domains as grazing incidence is approached.  Indeed, we see
that at normal incidence (blue curves) convergence to several digits
is achieved by using windows for which the length $(A-1)$ of each of
the two included flat windowed regions around the bump is of the order
of $8\lambda_s$ to $32\lambda_s$---a computational requirement which,
as shown in Section~\ref{sec:3.2}, can be greatly reduced. For
$\theta^{inc}=\pi/4$ (green curves) significantly worse accuracies are
obtained for each value of $\theta^{inc}$. As $\theta^{inc}$
approaches $\pi/2$ (red curves) the accuracy deteriorates much
further.

As noted in~\cite{BLPT16}, this difficulty can be explained by
consideration of certain arguments concerning bouncing geometrical
optics rays and the method of stationary phase. As shown in the
following section, the convergence as $A$ grows can be significantly
improved for all incidence angles. And, in fact, fast uniform
convergence for all incident angles, however close to grazing, can be
achieved.

% (Per Remark~\ref{remark_exact} we have $\theta_s\rightarrow\pm\pi/6$
% as $\theta^{inc}\rightarrow \pm\pi/2$ for $\lambda=2$, and
% $\theta_s\rightarrow \pm\pi/2$ as $\theta^{inc}\rightarrow \pm\pi/2$
% for $\lambda\rightarrow -\mu$.)

%\begin{table}[htb]
%\caption{Relative errors of the total field generated from preliminary WGF method for the Dirichlet problem of scattering by a semi-circular bump. {\color{red} Figure 3.2 in~\cite{BLPT16} seems much more dramatic than the experiments in this table. Let's discuss this.}}
%\centering
%\begin{tabular}{|c|c|c|c|c|c|c|}
%\hline
%\multirow{2}{*}{$(A-1)/\lambda_s$} & \multicolumn{3}{|c|}{$\lambda=2$}  & \multicolumn{3}{|c|}{$\lambda=-0.99$}\\
%\cline{2-7}
% & $\theta^{inc}=0$ & $\theta^{inc}=\frac{\pi}{4}$ & $\theta^{inc}=\frac{63\pi}{128}$  & $\theta^{inc}=0$ & $\theta^{inc}=\frac{\pi}{4}$ & $\theta^{inc}=\frac{63\pi}{128}$ \\
%\hline
%1 & 4.64E-1 & 7.73E-1 & 7.37E-1 & 5.24E-1 & 7.95E-1 & 7.43E-1  \\
%2 & 3.40E-1 & 6.72E-1 & 7.01E-1 & 3.56E-1 & 7.53E-1 & 7.85E-1  \\
%4 & 1.96E-1 & 3.79E-1 & 8.37E-1 & 1.08E-1 & 6.54E-1 & 8.98E-1  \\
%8 & 3.52E-2 & 1.80E-1 & 1.03E+0 & 1.07E-2 & 2.54E-1 & 8.81E-1  \\
%16 & 4.19E-3 & 5.66E-2 & 5.62E-1 & 8.89E-4 & 4.18E-2 & 7.65E-1  \\
%32 & 1.23E-4 & 7.20E-3 & 6.01E-2 & 3.86E-5 & 3.56E-3 & 3.08E-1  \\
%\hline
%\end{tabular}
%\label{Table3.1.1}
%\end{table}

\subsection{Uniformly accurate ``corrected'' formulation for all
  incidence angles}
\label{sec:3.2}

Utilizing the windowing function $\widetilde{w}_A$, equation
(\ref{DBIE}) may be re-expressed in the form \be
\label{ExactWDBIE1}
-\frac{1}{2}\phi+K'[\widetilde{w}_A\phi]=-Tu^{inc}-K'[(1-\widetilde{w}_A)\phi]
\quad\mbox{on}\quad\Gamma. \en An argument based on
integration-by-parts and stationary-phase presented in~\cite{BLPT16}
shows that for any positive integer $m$ there exists a constant $C_m$
independent of $A$, such that both the right-hand side term
$K'[(1-\widetilde{w}_A)\phi]$ and the windowing approximation error
$|\phi-\phi^*|$ (which results as that right-hand side term is
neglected, as in~\eqref{WDBIE1}) are smaller than $C_mA^{-m}$ as
$A\rightarrow\infty$, uniformly throughout the center region
$\{\widetilde{w}_A=1\}$ of the surface $\Gamma_A$.  However these
errors are not uniform with respect to the incidence angle: larger and
larger window sizes $A$ are required to correctly account for all
fields reflected and refracted by the planar surface as the incidence
angles are closer and and closer to grazing (i.e., as $\theta^{inc}$
approaches $\pm\pi/2$).

As proposed in \cite{BLPT16,P16} for acoustic layer scattering
problems, here we substitute the previously neglected right-hand side
term in~\eqref{ExactWDBIE1} by the expression that results as the
density $\phi$ is corrected, that is, it is replaced by the
corresponding ``flat-layer'' density
% solution, that is, by the corresponding traction of
% exact total field on $\Pi$, i.e.,
$\phi_\Pi =
(Tu_f^{tot})|_\Pi=(Tu_f^\mathrm{scat})|_\Pi+(Tu^{inc})|_\Pi$ that is
obtained for the problems of scattering by the flat surface $\Pi$. We
thus obtain the equation \be
\label{WDBIE2}
-\frac{1}{2}\phi^w+K'[\widetilde{w}_A\phi^w]=-Tu^{inc}-
K'[(1-\widetilde{w}_A)(Tu_f^{tot})|_\Pi] \quad\mbox{on}\quad\Gamma_A.
\en for the new approximate solution $\phi^w$. A superalgebraically
small portion of the field reflected by the windowed region reflects
back into the windowed region upon reflection from the plane outside
the windowed region. As a result, the substitution results in
superalgebraically small errors $|\phi-\phi^w|$ throughout the region
$\{w_A=1\}$.

In order to evaluate the right-hand term
$K'[(1-\widetilde{w}_A)(Tu_f^{tot})|_\Pi]$, which is given by an
integral over an unbounded domain, we note that
$(1-\widetilde{w}_A)(Tu_f^{tot})|_\Pi$ vanishes at all points at which
$\Gamma_A$ deviates from the flat surface $\Pi$. It follows that \ben
K'[(1-\widetilde{w}_A)(Tu_f^{tot})|_\Pi]=K'_\Pi[(1-\widetilde{w}_A)(Tu_f^{tot})|_\Pi]
\quad\mbox{on}\quad\Pi, \enn where, letting now $\nu_x$ denote the
normal to $\Pi$, the operator $K'_\Pi$ is defined by \be\label{planar}
K'_\Pi[\phi](x)&=&\int_\Pi T(\pa_x,\nu_x)E(x,y)\phi(y)ds_y.  \en But,
clearly, $K'_\Pi[\widetilde{w}_A(Tu_f^{tot})|_\Pi]$ can be evaluated
by means of numerical integration over the bounded region
$\Pi_A=\{x\in\Pi:\widetilde{w}_A(x)\ne 0\}$, and, using Green's
theorem~\cite{CGK02}, a closed form expression for
$K'_\Pi[(Tu_f^{tot})|_\Pi]$ results, \ben
K'_\Pi[(Tu_f^{tot})|_\Pi]=\begin{cases} -Tu^{inc}+
  \frac{1}{2}Tu_f^{tot} & \mbox{on}\quad\Gamma\cap\Pi, \cr
  Tu_f^\mathrm{scat} & \mbox{on}\quad (\Gamma\backslash\Pi)\cap\R^d_+,
  \cr -Tu^{inc} & \mbox{on}\quad (\Gamma\backslash\Pi)\cap\R^d_-,
\end{cases}
\enn ---and, therefore, the integral in~\eqref{planar} can be easily
be produced as a difference between these two quantities.

\begin{figure}[ht]
\centering
\includegraphics[scale=0.3]{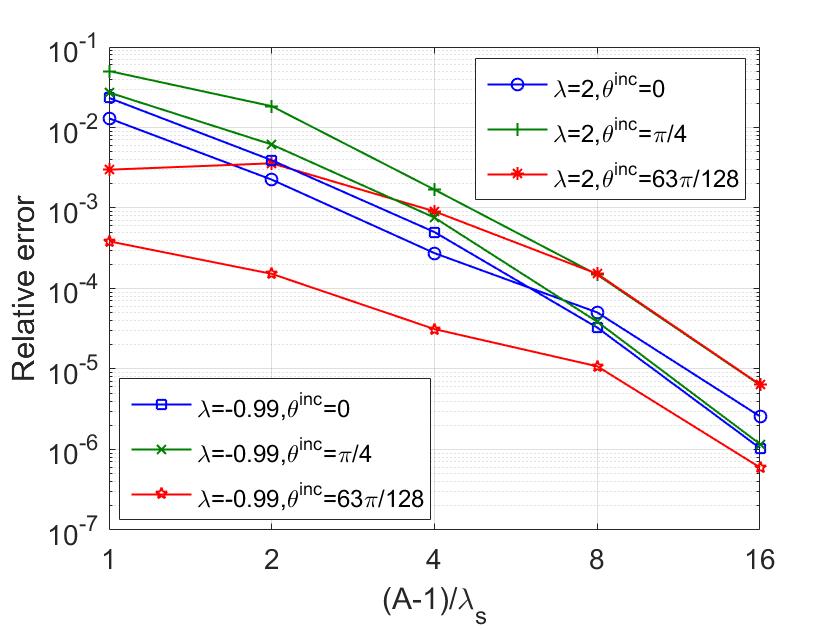}
\caption{Relative errors $\epsilon_\infty$ in the total field
  resulting from the uniformly-accurate corrected WGF method for the
  Dirichlet problem of scattering by a semi-circular bump.}
\label{Table3.1.2}
\end{figure}

For the evaluation of the near-field, we follow \cite{BLPT16} and
substitute $\phi$ in the representation \ben
u^\mathrm{scat}(x)=\int_\Gamma E(x,y)\phi(y)ds_y \enn by
$w_A\phi^w+(1-w_A)(Tu_f^{tot}|_\Pi)$ which yields \be
\label{nearD}
u^\mathrm{scat}(x)&=& \int_{\Gamma_A} E(x,y)w_A(y)\phi^w(y)ds_y- \int_{\Pi} E(x,y)w_A(y)Tu_f^{tot}(y)ds_y\nonumber\\
&\quad& +\begin{cases}
u_f^\mathrm{scat} & \mbox{in}\quad \Omega\cap\R^d_+,\cr
-u^{inc} & \mbox{in}\quad \Omega\cap\R^d_-.
\end{cases}
%&\quad& +\int_{\Pi} E(x,y)Tu_f^{tot}(y)ds_y
\en The character of the overall approach is demonstrated in
Figure~\ref{Table3.1.2}, which presents the the relative errors in the
total field $u^{tot}$ on the line segment
$\{x\in\R^2: -1\le x_1\le 1, x_2=2\}$ (which were evaluated by
comparison with a WGF solution with $(A - 1)/\lambda_s =32$). Comparison
with the results of the preliminary WGF method demonstrated in
Figure~\ref{Table3.1.1} demonstrates the improvements provided by the
present uniformly-accurate algorithm: much faster convergence which,
as desired, is uniform for all incident angles; additional numerical
illustrations of the character of the algorithm are presented in
Section \ref{sec:5}.

%\begin{table}[htb]
%\caption{Relative errors of the total field generated from the modified WGF method for the Dirichlet problem of scattering by a semi-circular bump.}
%\centering
%\begin{tabular}{|c|c|c|c|c|c|c|c|}
%\hline
%\multirow{2}{*}{$\omega$} & \multirow{2}{*}{$(A-1)/\lambda_s$} & \multicolumn{3}{|c|}{$\lambda=2$}  & \multicolumn{3}{|c|}{$\lambda=-0.99$} \\
%\cline{3-8}
%& & $\theta^{inc}=0$ & $\theta^{inc}=\frac{\pi}{4}$ & $\theta^{inc}=\frac{63\pi}{128}$  & $\theta^{inc}=0$ & $\theta^{inc}=\frac{\pi}{4}$ & $\theta^{inc}=\frac{63\pi}{128}$ \\
%\hline
% & 1 & 1.64E-2 & 2.09E-2 & 2.05E-2 & 6.75E-3 & 5.35E-3 & 2.54E-3  \\
% & 2 & 8.49E-3 & 8.02E-3 & 9.96E-3 & 7.73E-4 & 8.79E-4 & 4.28E-4  \\
%4 & 4 & 4.92E-4 & 5.36E-4 & 5.16E-4 & 7.74E-5 & 7.42E-5 & 3.40E-5  \\
% & 8 & 7.76E-5 & 6.04E-5 & 6.10E-5 & 2.37E-6 & 2.38E-6 & 1.50E-6  \\
% & 16 & 1.55E-6 & 9.80E-7 & 1.10E-6 & 9.11E-8 & 2.82E-7 & 4.46E-7  \\
%\hline
% & 1 & 1.30E-2 & 5.03E-2 & 2.99E-3 & 2.33E-2 & 2.72E-2 & 3.86E-4  \\
% & 2 & 2.25E-3 & 1.84E-2 & 3.59E-3 & 3.92E-3 & 6.15E-3 & 1.52E-4  \\
%20 & 4 & 2.75E-4 & 1.69E-3 & 9.08E-4 & 5.02E-4 & 7.59E-4 & 3.13E-5  \\
% & 8 & 5.04E-5 & 1.48E-4 & 1.52E-4 & 3.27E-5 & 3.86E-5 & 1.08E-5  \\
% & 16 & 2.58E-6 & 6.34E-6 & 6.42E-6 & 1.04E-6 & 1.17E-6 & 5.98E-7  \\
%\hline
%\end{tabular}
%\label{Table3.1.2}
%\end{table}

\begin{figure}[ht]
\centering
\begin{tabular}{ccc}
\includegraphics[scale=0.15]{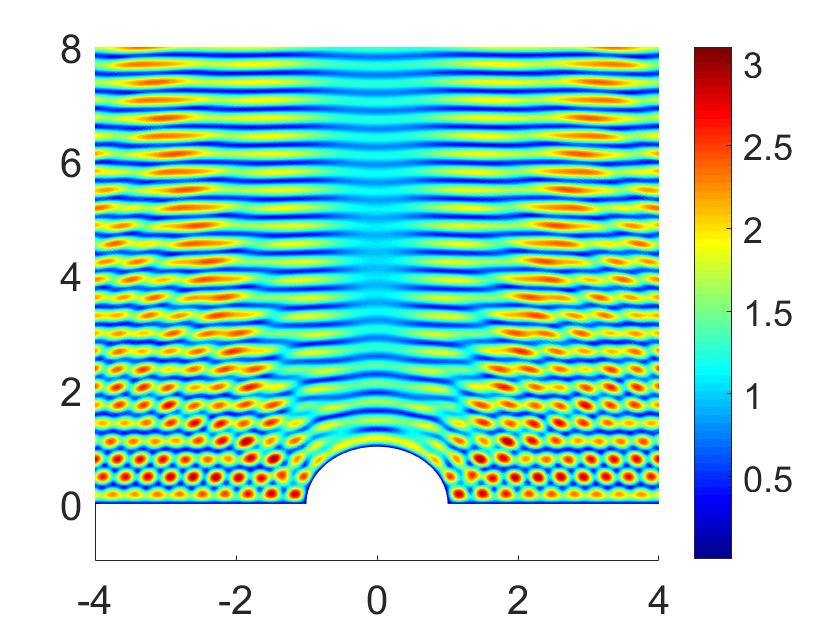} &
\includegraphics[scale=0.15]{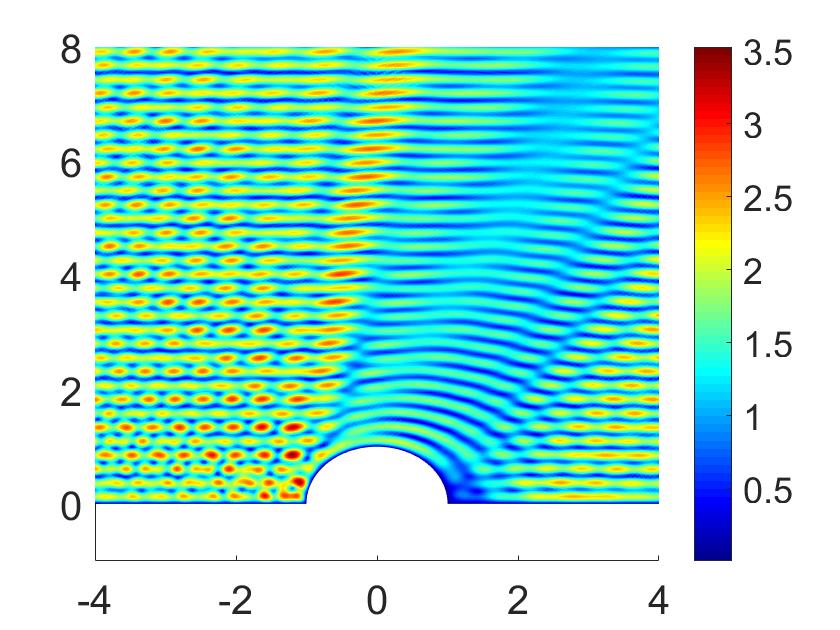} &
\includegraphics[scale=0.15]{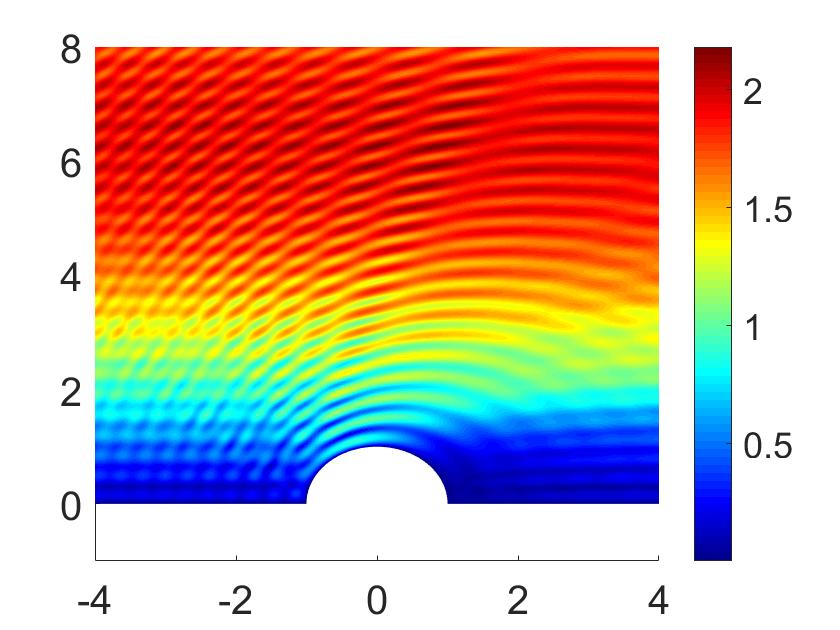} \\
(a) $\theta^{inc}=0$ & (b) $\theta^{inc}=\frac{\pi}{4}$ & (c) $\theta^{inc}=\frac{63\pi}{128}$ \\
\end{tabular}
\caption{Absolute values of the total field  resulting from the WGF method for the Dirichlet problem of scattering of a plane pressure wave by a semi-circular bump where $\omega=20$, $\lambda=2$, $A=1+16\lambda_s$.}
\label{Figure3.1.1}
\end{figure}

\begin{remark}
  A version of the windowed formulation of the integral equation
  (\ref{NBIE}) suitable for treatment of the Neumann problem can similarly be obtained. The resulting integral equation reads
\be
\label{WNBIE}
\frac{1}{2}\psi^w+K[\widetilde{w}_A\psi^w]=u^{inc}+K_\Pi[(\widetilde{w}_A-1)(u_f^{tot})|_\Pi] \quad\mbox{on}\quad \Gamma_A,
\en
where the operator $K_\Pi$ is defined by
\ben
K_\Pi[\psi](x)&=&\int_\Pi (T(\pa_y,\nu_y)E(x,y))^\top\psi(y)ds_y.
\enn
The term $K_\Pi[\widetilde{w}_A(u_f^{tot})|_\Pi]$ can be evaluated by means of numerical integration over the bounded region $\Pi_A$ and the expression $K_\Pi[(u_f^{tot})|_\Pi]$ can be computed in closed form:
\ben
K_\Pi[(u_f^{tot})|_\Pi]=\begin{cases}
u^{inc}- \frac{1}{2}u_f^{tot} & \mbox{on}\quad\Gamma\cap\Pi, \cr
u^{inc}- u_f^{tot} & \mbox{on}\quad (\Gamma\backslash\Pi)\cap\R^d_+, \cr
u^{inc} & \mbox{on}\quad (\Gamma\backslash\Pi)\cap\R^d_-.
\end{cases}
\enn Furthermore, substituting
$\psi = w_A\psi^w+(1-w_A)u_f^{tot}|_\Pi$ in the scattered field
representation \ben u^\mathrm{scat}(x)=-\int_\Gamma
(T(\pa_y,\nu_y)E(x,y))^\top\widetilde{w}_A\psi(y)ds_y \enn for the
Neumann problem yields \be
\label{nearN}
u^\mathrm{scat}(x)&=&-\int_\Gamma (T(\pa_y,\nu_y)E(x,y))^\top\widetilde{w}_A\psi^w(y)ds_y+ \int_\Pi (T(\pa_y,\nu_y)E(x,y))^\top \widetilde{w}_Au_f^{tot}(y)ds_y\nonumber\\
&\quad& +\begin{cases}
u_f^\mathrm{scat} & \mbox{in}\quad \Omega\cap\R^d_+,\cr
-u^{inc} & \mbox{in}\quad \Omega\cap\R^d_-.
\end{cases}
\en
for the evaluation of near-field.
\end{remark}

\begin{remark}
  The expressions (\ref{nearD}) and (\ref{nearN}) generally do not
  provide accurate approximations of either far-fields or near fields
  outside bounded subsets of $[-cA,cA]\times \R^{d-1}$. This
  difficulty can be tackled~\cite{BLPT16} via an application of the
  Green theorem on a curve $S$ contained in $[-cA,cA]\times \{x_d>0\}$
  and surrounding the defect, together with the layer Green
  function-based method discussed in Section~\ref{sec:2.3}---which,
  for such near- and far-field cases, for which the source and
  observation points are at a large or even infinite distances from
  each other, the layer Green function can be obtained rapidly.  
\end{remark}

\section{Numerical implementation}
\label{sec:4}

The iterative solvers for solution of the discrete versions of
(\ref{WDBIE2}) and (\ref{WNBIE}) rely on the numerical evaluation of
integral operators and the iterative linear algebra solver GMRES. This
section presents the 2D algorithms for the numerical evaluation, for a
given density $\psi$, of the quantities $K[\widetilde{w}_A\psi]$,
$K'[\widetilde{w}_A\psi]$, $K_\Pi[\widetilde{w}_A\psi]$ and
$K_\Pi'[\widetilde{w}_A\psi]$ associated with the WGF method for
the solution of the Dirichlet and Neumann problems. For the numerical
implementation in 3D, in turn, we utilize the methods presented
in~\cite{BG18,BY19}.

\subsection{Reformulation in terms of composite differential/weakly-singular operators}
\label{sec:4.1}

As discussed in Section~\ref{sec:1}, the methods~\cite{YHX17} can be used to express the
quantities \be
\label{KWA}
K[\widetilde{w}_A\psi](x)&=&\int_{\Gamma_A} (T(\pa_y,\nu_y)E(x,y))^\top \widetilde{w}_A(y)\psi(y)ds_y,\quad x\in\Gamma_A,\\
\label{KPWA}
K'[\widetilde{w}_A\psi](x)&=&\int_{\Gamma_A}
T(\pa_x,\nu_x)E(x,y)\widetilde{w}_A(y)\psi(y)ds_y,\quad x\in\Gamma_A,
\en 
in the forms
\be
\label{operatorK}
K[\widetilde{w}_A\psi](x) &=& K_1[\widetilde{w}_A\psi](x)+K_2\left[ T_0(\widetilde{w}_A\psi) \right](x),\\
\label{operatorKP}
K'[\widetilde{w}_A\psi](x) &=& K_1'[\widetilde{w}_A\psi](x)+T_0K'_2\left[\widetilde{w}_A\psi \right](x),
\en
where the operators $K_1,K_2,K'_1,K'_2$ are given by
\ben
K_1[\phi](x) &=& \int_{\Gamma_A} H_1(x,y)\phi(y)ds_y, \quad H_1(x,y)=\frac{\pa \gamma_{k_s}(x,y)}{\pa\nu_y}I- \nabla_y[\gamma_{k_s}(x,y)-\gamma_{k_p}(x,y)]\nu_y^\top,\\
K_2[\phi](x) &=& \int_{\Gamma_A} H_2(x,y)\phi(y)ds_y, \quad H_2(x,y)=[2\mu E(x,y)-\gamma_{k_s}(x,y)I]\begin{pmatrix}
0 & -1 \\
1 & 0
\end{pmatrix},\\
K_1'[\phi](x) &=& \int_{\Gamma_A} H_3(x,y)\phi(y)ds_y, \quad H_3(x,y)=\frac{\pa \gamma_{k_s}(x,y)}{\pa\nu_x}I- \nu_x\nabla_y^\top[\gamma_{k_s}(x,y)-\gamma_{k_p}(x,y)],\\
K_2'[\phi](x) &=& \int_{\Gamma_A} H_4(x,y)\phi(y)ds_y, \quad H_4(x,y)=\begin{pmatrix}
0 & -1 \\
1 & 0
\end{pmatrix}[2\mu E(x,y)-\gamma_{k_s}(x,y)I],\\
\enn and where
\[T_0=\nu^\perp\cdot\nabla\]
denotes the tangential
derivative. The quantities $K_\Pi[\widetilde{w}_A\psi]$ and
$K_\Pi'[\widetilde{w}_A\psi]$ can be re-expressed in a similar
manner. In view of these reformulations, the integral operators
introduced in Section \ref{sec:3.2} can be evaluated numerically as a
sum of compositions involving the numerical differentiation operator
$T_0$ as well as integral operators of the form \be
\label{objintegral}
\mathcal{H}[\psi](x)= \int_{\Gamma_0} H(x,y)\psi(y)ds_y,\quad
\Gamma_0=\Gamma_A\quad\mbox{or}\quad \Pi_A, \en in which the kernel
$H(x,y)$ is only weakly singular. The remainder of this section
presents the algorithms we propose for numerical evaluation of
operators of these two types, including a two dimensional version of
the rectangular-polar Chebyshev-based quadrature method~\cite{BG18}
for weakly singular operators of the form (\ref{objintegral}) and
Chebyshev-based differentiation algorithms.

\subsection{Surface decomposition and discretization}
\label{sec:4.2}

The proposed algorithm evaluates weakly singular integrals of the
form~\eqref{objintegral} by first partitioning $\Gamma_0$ into a
finite number $M$ of parametrized patches $\Gamma_q$, $q=1,\dots,M$:
\ben \Gamma_0=\bigcup_{q=1}^M\Gamma_q,\quad \Gamma_q=\{x^q(t): [-1,1]
\rightarrow \R^2\}.  \enn (It is assumed that each corner point
$x\in \Gamma_0$, if any such point exists, is located at
parametrization endpoints $t=1$ or $-1$ of the patches $\Gamma_q$ that
contain $x$.)  Clearly, then, the integral (\ref{objintegral}) can be
expressed as a sum of integrals over each of the patches: \ben
\mathcal{H}(x)= \sum_{q=1}^M \mathcal{H}_q(x),\quad \mathcal{H}_q(x)=
\int_{\Gamma_q} H(x,y)\phi(y)ds_y.  \enn Using the parametrization
$x = x^q(t)$ for the patch $\Gamma_q$ we obtain \be
\label{objintegral1}
\mathcal{H}_q(x)= \int_{-1}^1 \widetilde{H}(x,t)\widetilde{\phi}(t)J^q(t)dt,
\en
where $\widetilde{H}(x,t)=H(x,x^q(t))$, $\widetilde{\phi}(t)=\phi(x^q(t)$ and $J^q(t)=|dx^q(t)/dt|)\ne 0$ denotes the surface Jacobian.

To treat the singular character of integral-equation densities at
corners in a general and robust manner, we introduce a change of
variables~\cite{BG18} on the parametrization variables $t$, a number
of whose derivatives vanish at the corners. In detail, defining the
function \ben
w_p(\tau)=2\pi\frac{[v_p(\tau)]^p}{[v_p(\tau)]^p+[v_p(2\pi-\tau)]^p},\quad
0\le\tau\le2\pi, \enn where \ben
v_p(\tau)=\left(\frac{1}{p}-\frac{1}{2}\right)\left(\frac{\pi-\tau}{\pi}\right)^3
+\frac{1}{p}\left(\frac{\tau-\pi}{\pi}\right) +\frac{1}{2}, \enn
(whose derivatives vanish up to order $p-1$ at the endpoints $\tau=0$
and $\tau=2\pi$), we define the smoothing change-of-variables
$t=\eta_t^q(\tau)$ for the patch $\Gamma_q$ according to the 
expressions 
\be
\label{COV}
t=\eta_t^q(\tau)=\begin{cases}
\tau, & \mbox{No corner at either $\Gamma_q$ endpoint}, \cr
-1+ \frac{1}{\pi}w_p(\pi(\tau+1)), & \mbox{Corners at both $\Gamma_q$ endpoints}, \cr
-1+ \frac{2}{\pi}w_p(\pi(\tau+1)/2), & \mbox{Corner at the $t=-1$ $\Gamma_q$ endpoint only}, \cr
-3+ \frac{2}{\pi}w_p(\pi+\pi(\tau+1)/2), & \mbox{Corner at the $t=1$ $\Gamma_q$ endpoint only}.
\end{cases}
\en
Incorporating the change of variables (\ref{COV}), we obtain
\be
\label{objintegral2}
\mathcal{H}_q(x)= \int_{-1}^1 \widetilde{H}(x,\eta_t^q(\tau))\widetilde{\phi}(\eta_t^q(\tau)) J^q(\eta_t^q(\tau)) \frac{d\eta_t^q(\tau)}{d\tau}d\tau,
\en

Considering the distance \ben
\mbox{dist}_{x,\Gamma_q}:=\min_{u\in[-1,1]} \left\{|x-x^q(u)|
\right\}, \enn between the point $x$ and the patch $\Gamma_q$, a
number of ``singular'', ``near-singular'' and ``regular'' integration
problems arise as described in Section \ref{sec:4.3}. For accuracy and
efficiency our algorithm evaluates these integrals are produced by
means of Fej\'er's first quadrature rule, which effectively exploits
the discrete orthogonality property satisfied by the Chebyshev
polynomials in the Chebyshev meshes. Denoting by $u_j\in [-1,1]$
($j=0,\cdots,N-1$) the $N$ Chebyshev points \ben
u_j=\cos\left(\frac{2j+1}{2N}\pi\right),\quad j=0,\cdots,N-1, \enn we
utilize discretization points in each patch $\Gamma_q$ according to
$x_i^q=x^q(\eta_t^q(u_i)),\quad i=0,\cdots,N-1$. Then, a given density
$\varphi$ with values $\varphi_i^q = \varphi(x_i^q)$ is approximated
by means of the Chebyshev expansion \ben \varphi(x)\approx
\sum_{i=0}^{N-1} \varphi_i^q a_i(u),\quad x\in\Gamma_q, \enn where
the quantities
\ben a_i(u)= \frac{1}{N}\sum_{n=0}^{N-1}\alpha_n T_n(u_i)T_n(u),\quad
\alpha_n=\begin{cases} 1, & n=0,\cr 2, & n\ne 0.
\end{cases}
\enn satisfy the discrete-orthogonality relations \ben
a_i(u_n)=\begin{cases} 1, & n=i,\cr 0, & \mathrm{otherwise}.
\end{cases}
\enn
\begin{figure}[htbp]
\centering
\begin{tabular}{ccc}
\includegraphics[width=4.5cm,height=4.5cm]{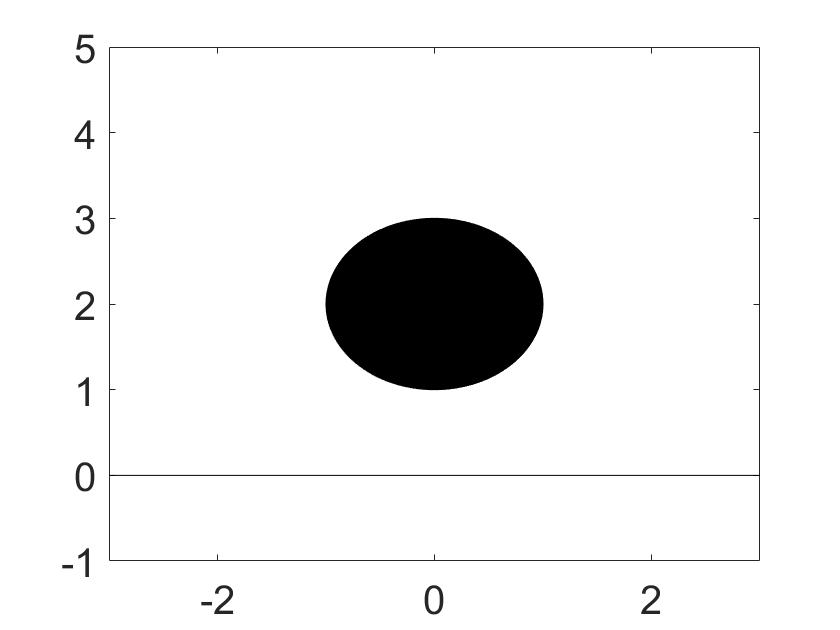} &
\includegraphics[width=4.5cm,height=4.5cm]{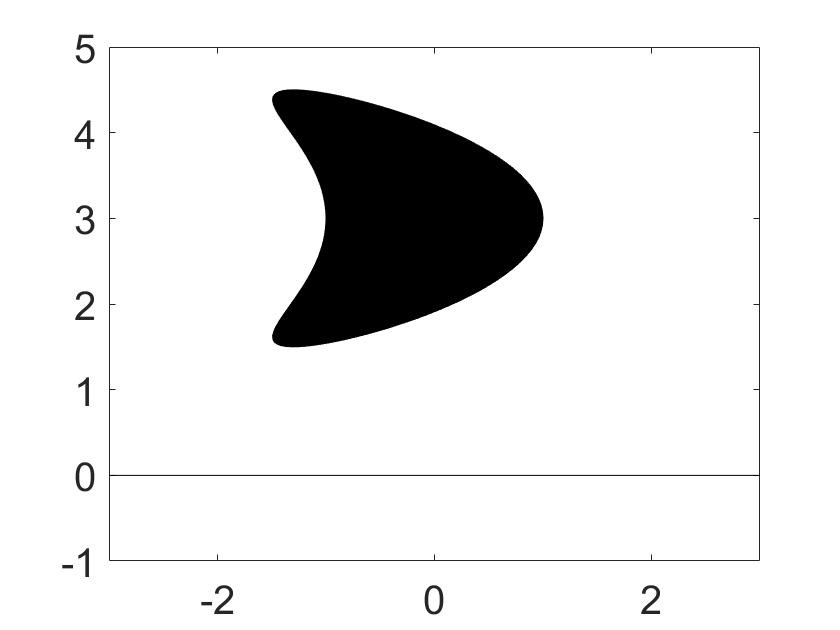} &
\includegraphics[width=4.5cm,height=4.5cm]{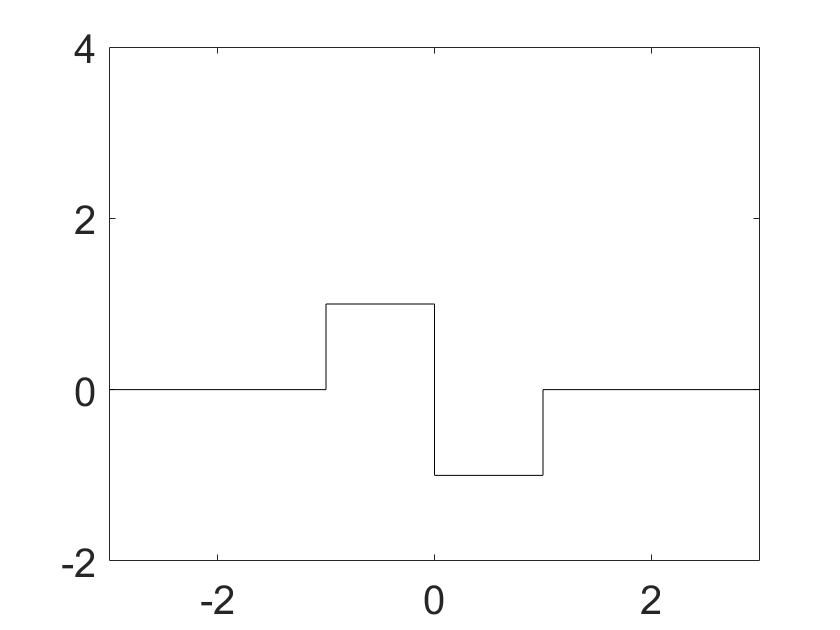} \\
(a) disc within half-space & (b) kite within half-space& (c) local boundary perturbation\\
\end{tabular}
\caption{Various 2D half-space considered in this paper.}
\label{Figure2D.1}
\end{figure}

\subsection{Non-adjacent and adjacent integration}
\label{sec:4.3}

Let $x$ be one of the discretization points on $\Gamma_A$. In the
"non-adjacent" integration case, in which the point $x$ is far from
the integration patch (i.e., $\mbox{dist}_{x,\Gamma_q}> \tau$ for some
tolerance $\tau>0$), the integrand $\mathcal{H}_q(x)$ is smooth, and
the integral over $\Gamma_q$ can be accurately evaluated by means of
Fej\'er's first quadrature rule \be
\label{nonadjencent}
\mathcal{H}_q(x) &\approx& \sum_{n=0}^{N-1}
w_n\widetilde{H}(x,\eta_t^q(u_n)) \varphi_{n}^q
J^q(\eta_t^q(u_n))\left(\frac{d\eta_t^q(\tau)}{d\tau}\Big|_{\tau=u_n}\right), \en where $w_j,j=0,\cdots,N-1$ are the quadrature weights \ben
w_j=\frac{2}{N}\left(1-2\sum_{l=1}^{\lfloor N/2\rfloor}
  \frac{1}{4l^2-1}\cos(lu_j)\right),\quad j=0,\cdots,N-1.  \enn

In the "adjacent" integration case, in which the point $x$ either lies
within the integration patch or is "close" to it (i.e.,
$\mbox{dist}_{x,\Gamma_q}\le \tau$), the problem of evaluation of
$\mathcal{H}_q(x)$ presents a challenge in view of the singularity or
nearly-singularity of its kernel. To tackle this difficulty we apply a
change of variables whose derivatives vanish at the singularity or,
for nearly singular problems, at the point in the integration patch
that is closest to the singularity---in either case, the coordinates
$u^q\in [-1,1]$ of the point around which refinements are performed
are given by \ben \widetilde{u}^q={\arg\min}_{u\in[-1,1]}
\left\{|x-x^q(\eta_t^q(u))| \right\}.  \enn The quantities
$\widetilde{u}^q$ can be found by means of an appropriate minimization
algorithm such as the golden section search algorithm. Making use of
the mapping $w_p$ defined in Section \ref{sec:4.2} we construct the
change of variables \ben \xi_\alpha(t)=\begin{cases}
  \alpha+\frac{\mbox{sgn}(t)-\alpha}{\pi}w_p(\pi|t|), & \alpha\ne
  \pm1, \cr
  \alpha-\frac{1+\alpha}{\pi}w_p\left(\pi\frac{|t-1|}{2}\right), &
  \alpha=1, \cr
  \alpha+\frac{1-\alpha}{\pi}w_p\left(\pi\frac{|t+1|}{2}\right), &
  \alpha=-1,
\end{cases}
\enn where $\mbox{sgn}(x)$ equals $1$, $-1$ or $0$ according to
whether $x>0$, $x<0$ or $x=0$, respectively. Applying the Chebyshev
expansion of the density $\varphi$, the above change of variables and
the Fej\'er's first quadrature rule, we obtain \be
\label{adjencent}
\mathcal{H}_q(x) &\approx& \sum_{n=0}^{N-1}\varphi_{n}^q \sum_{m=0}^{N^\beta-1} \widetilde{w}_m\widetilde{H}(x,s^q_m) \widetilde{J}^q(s^q_m) \widetilde{a}_n(s^q_m) \xi_{\widetilde{u}^q}'(\widetilde{u}_m) \left(\frac{d\eta_t^q(\tau)}{d\tau}\Big|_{\tau=\xi_{\widetilde{u}^q}(\widetilde{u}_m)}\right)
\en
where
\ben
s^q_m=\eta_t^q(\xi_{\widetilde{u}^q}(\widetilde{u}_m)),
\enn
and where  the quadrature nodes and weights are given by
\ben
\widetilde{u}_j=\cos\left(\frac{2j+1}{2N^\beta}\pi\right),\quad j=0,\cdots,N^\beta-1,
\enn
and
\ben
\widetilde{w}_j=\frac{2}{N^\beta}\left(1-2\sum_{l=1}^{\lfloor N^\beta/2\rfloor} \frac{1}{4l^2-1}\cos(l\widetilde{u}_j)\right),\quad j=0,\cdots,N^\beta-1,
\enn
respectively. Using sufficiently large numbers $N^\beta$ of discretization points to accurately resolve the challenging integrands, all singular and nearly singular problems can be treated with high accuracy under discretizations that are not excessively fine.

\begin{table}[htbp]
\centering
\begin{tabular}{|c|c|c|c|c|c|c|c|}
\hline
\multirow{2}{*}{$\omega$} & \multirow{2}{*}{$A/\lambda_s$} & \multicolumn{3}{|c|}{Disc-shaped}  & \multicolumn{3}{|c|}{Kite-shaped}\\
\cline{3-8}
& & $\theta^{inc}=0$ & $\theta^{inc}=\frac{\pi}{4}$ & $\theta^{inc}=\frac{63\pi}{128}$
& $\theta^{inc}=0$ & $\theta^{inc}=\frac{\pi}{4}$ & $\theta^{inc}=\frac{63\pi}{128}$  \\
\hline
  & 2  & 1.54E-2 & 2.61E-2 & 1.18E-2 & 3.79E-2 & 5.13E-2 & 8.76E-3 \\
  & 4  & 1.62E-3 & 8.10E-3 & 3.92E-3 & 3.29E-3 & 1.46E-2 & 8.92E-3 \\
4 & 8  & 1.49E-4 & 5.20E-4 & 2.73E-4 & 1.48E-4 & 3.60E-4 & 5.58E-4 \\
  & 16 & 2.12E-6 & 2.49E-6 & 1.14E-6 & 1.85E-6 & 1.30E-6 & 4.43E-6 \\
  & 32 & 2.99E-8 & 4.29E-8 & 1.98E-8 & 3.27E-8 & 2.06E-8 & 1.11E-7 \\
\hline
   & 2  & 1.98E0  & 1.35E0  & 2.71E-1 & 8.34E-1 & 1.65E0  & 1.06E0  \\
   & 4  & 4.92E-2 & 2.81E-2 & 2.25E-2 & 5.30E-1 & 1.54E0  & 1.33E0  \\
20 & 8  & 4.72E-3 & 1.09E-2 & 2.85E-3 & 7.80E-3 & 1.24E-2 & 1.63E-3 \\
   & 16 & 3.22E-5 & 1.45E-4 & 6.90E-5 & 7.17E-5 & 2.13E-4 & 2.93E-4 \\
   & 32 & 9.55E-7 & 1.46E-6 & 1.81E-6 & 9.30E-7 & 2.12E-6 & 1.53E-6 \\
\hline
\end{tabular}
\caption{Relative errors $\epsilon_\infty$ in the total field  resulting from the  WGF method for the Dirichlet problems of scattering by a disc-shaped and a kite-shaped obstacle within a half space.}
\label{Table2D.1}
\end{table}

\begin{table}[htbp]
\centering
\begin{tabular}{|c|c|c|c|c|c|c|c|}
\hline
\multirow{2}{*}{$\omega$} & \multirow{2}{*}{$A/\lambda_s$} & \multicolumn{3}{|c|}{Disc-shaped}  & \multicolumn{3}{|c|}{Kite-shaped}\\
\cline{3-8}
& & $\theta^{inc}=0$ & $\theta^{inc}=\frac{\pi}{4}$ & $\theta^{inc}=\frac{63\pi}{128}$
& $\theta^{inc}=0$ & $\theta^{inc}=\frac{\pi}{4}$ & $\theta^{inc}=\frac{63\pi}{128}$  \\
\hline
  & 2  & 4.31E-2 & 2.95E-2 & 1.78E-2 & 5.85E-2 & 5.47E-2 & 4.05E-2 \\
  & 4  & 4.48E-3 & 7.24E-3 & 5.87E-3 & 7.75E-3 & 1.60E-2 & 8.51E-3 \\
4 & 8  & 3.87E-4 & 3.36E-4 & 2.60E-4 & 2.67E-4 & 6.06E-4 & 6.67E-4 \\
  & 16 & 8.16E-6 & 3.65E-6 & 2.83E-6 & 6.48E-6 & 7.84E-6 & 8.37E-6 \\
  & 32 & 1.31E-7 & 5.07E-8 & 1.51E-7 & 5.97E-8 & 1.04E-7 & 1.30E-7 \\
\hline
   & 2  & 2.02E0  & 2.24E0  & 2.95E-1 & 7.31E-1 & 1.50E0  & 1.39E0  \\
   & 4  & 1.66E-1 & 2.88E-1 & 1.91E-1 & 4.27E-1 & 1.55E0  & 1.41E0  \\
20 & 8  & 5.37E-3 & 2.15E-2 & 3.37E-3 & 6.40E-3 & 1.16E-2 & 4.98E-3 \\
   & 16 & 7.64E-5 & 1.67E-4 & 8.09E-5 & 9.05E-5 & 1.96E-4 & 5.73E-4 \\
   & 32 & 2.21E-6 & 3.18E-6 & 2.37E-6 & 1.01E-6 & 3.52E-6 & 2.38E-6 \\
\hline
\end{tabular}
\caption{Relative errors $\epsilon_\infty$ in the total field  resulting from the  WGF method for the Neumann problems of scattering by a disc-shaped and a kite-shaped obstacle within a half space.}
\label{Table2D.2}
\end{table}
\subsection{Evaluation of tangential derivatives}
\label{sec:4.4}

Finally, we describe the implementation we use for the evaluation of
the tangential derivative operator $T_0$. On each patch $\Gamma_q$,
applying the surface parametrization for a given density $\varphi$ we
have \ben
\varphi(x)=\phi(x^q(\eta_t^q(\tau)))=\phi(x^q(\eta_t^q(\cos\theta))),
\quad\theta\in[0,\pi].  \enn It follows the tangential derivative of
$\varphi$ on $\Gamma_q$ is given by \ben
(T_0\varphi)(x)=-\frac{1}{J^q(\eta_t^q(\cos\theta)) \sin\theta
  \left(\frac{d\eta_t^q(\tau)}{d\tau}\Big|_{\tau=\cos\theta}\right)}
\frac{d\varphi(x^q(\eta_t^q(\cos\theta)))}{d\theta}.  \enn (Note that
the tangential derivative operator is evaluated at the Chebyshev
points $\theta_j=\pi\frac{2j+1}{2N}$, $j=0,1,\cdots,N-1$, at which
$J^q(\eta_t^q(\cos\theta_j)) \sin\theta_j
\left(\frac{d\eta_t^q(\tau)}{d\tau}\Big|_{\tau=\cos\theta_j}\right)\ne
0$.) Given the values of density $\phi(x^q(\eta_t^q(\cos\theta)))$ at
the Chebyshev points $\theta_j=\pi\frac{2j+1}{2N}$,
$j=0,1,\cdots,N-1$, the necessary derivative with respect to $\theta$
can be evaluated by extending the density as an even function in
$[-\pi,\pi]$ and using FFT.

\begin{figure}[htbp]
\centering
\begin{tabular}{ccc}
\includegraphics[scale=0.15]{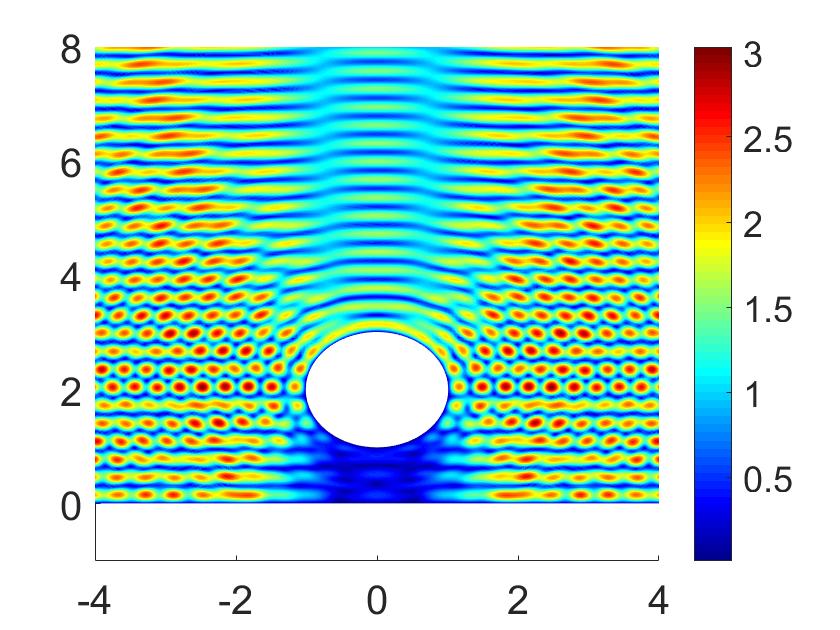} &
\includegraphics[scale=0.15]{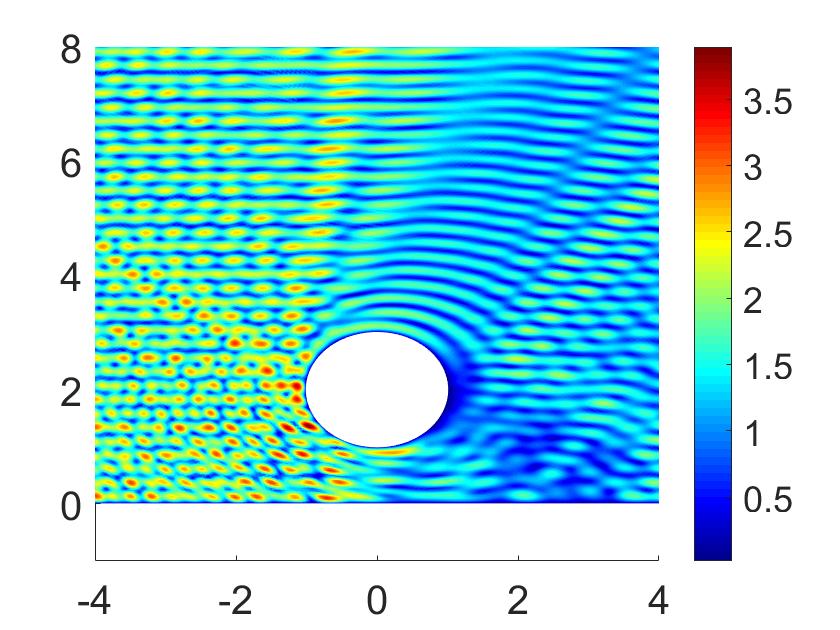} &
\includegraphics[scale=0.15]{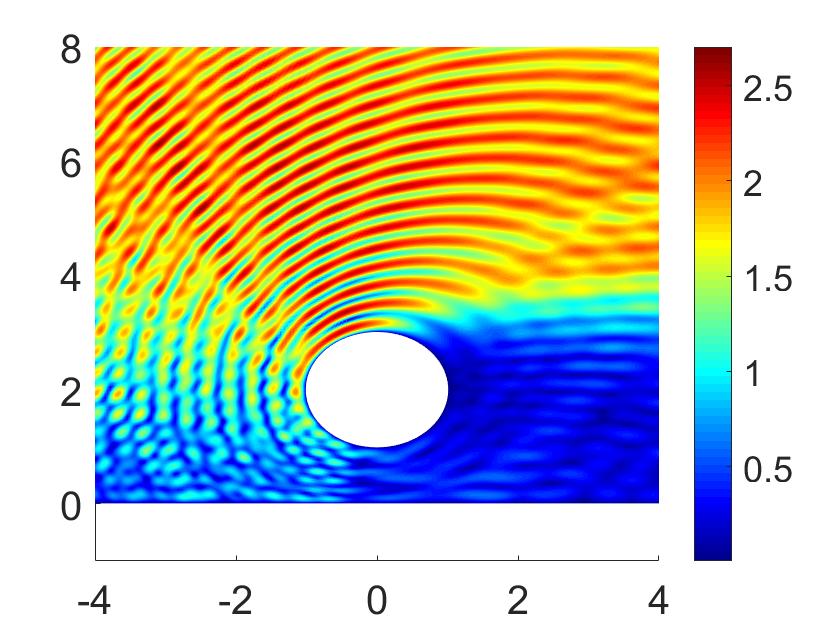} \\
(a) $\theta^{inc}=0$ & (b) $\theta^{inc}=\frac{\pi}{4}$ & (c) $\theta^{inc}=\frac{63\pi}{128}$ \\
\end{tabular}
\caption{Absolute values of the total field  resulting from the WGF method for the Dirichlet problem of scattering by a disc-shaped obstacle where $\omega=20$, $A=32\lambda_s$.}
\label{Figure2D.2}
\end{figure}

\begin{figure}[htbp]
\centering
\begin{tabular}{ccc}
\includegraphics[scale=0.15]{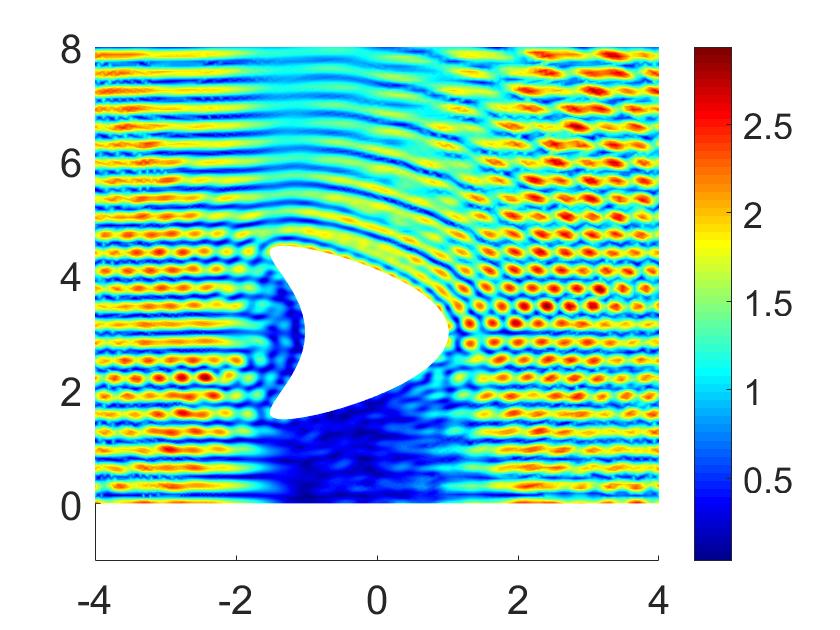} &
\includegraphics[scale=0.15]{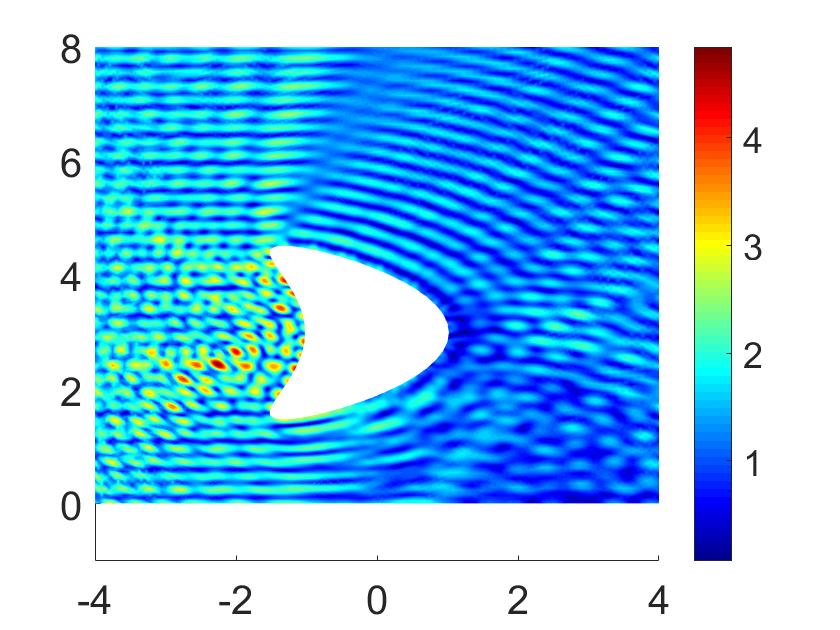} &
\includegraphics[scale=0.15]{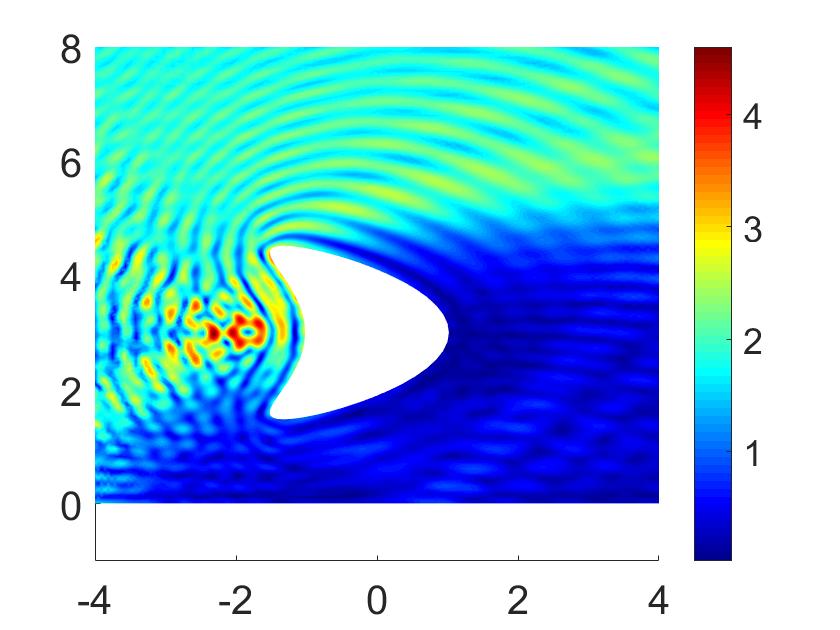} \\
(a) $\theta^{inc}=0$ & (b) $\theta^{inc}=\frac{\pi}{4}$ & (c) $\theta^{inc}=\frac{63\pi}{128}$ \\
\end{tabular}
\caption{Absolute values of the total field  resulting from the WGF method for the Neumann problem of scattering by a kite-shaped obstacle where $\omega=20$, $A=32\lambda_s$.}
\label{Figure2D.3}
\end{figure}

\section{Numerical results}
\label{sec:5}

The two-dimensional numerical results presented in section
\ref{sec:3}, and, particularly, Figures~\ref{Table3.1.2}
and~\ref{Figure3.1.1}, demonstrate the advantages inherent in the
uniformly-accurate fixed-windowed integral formulation~(\ref{WDBIE2}),
namely, fast convergence uniformly over all incidence angles. The
present section, in turn, presents a variety of additional numerical
examples in both 2D and 3D, which demonstrate the efficiency and
accuracy of the proposed WGF method. Solutions for the integral
equations were produced by means of the fully complex version of the
iterative solver GMRES. All of the numerical tests were obtained by
means of Fortran numerical implementations, parallelized using OpenMP,
on a single node (twenty-four computing cores) of a dual socket Dell
R420 with two Intel Xenon E5-2670 v3 2.3 GHz, 128GB of RAM. In all
cases, unless otherwise stated, the values $\lambda=2$, $\mu=1$,
$\rho=1$, $p=8$, $c=0.7$ were used and the relative errors reported
were calculated in accordance with the expression \be\label{rel_error}
\epsilon_\infty=\frac{\max_{x\in
    S}|u^{\mbox{num}}(x)-u^{\mbox{ref}}(x)|}{\max_{x\in
    S}|u^{\mbox{ref}}(x)|}, \en where $u^{\mbox{ref}}$ is produced by
means of numerical solution with a sufficiently fine discretization
and a sufficiently large value of $A$, and where $S$ is a suitably
selected line segment (2D) or square plane (3D) above the defect, and
at a distance from it no larger than 2.  The parameters $M$, $N$,
$N^\beta$ were selected in such a way that the errors arising from the
numerical integration are negligible in comparison with the
smooth-windowing errors. Throughout this section aEn denotes
$a\times10^{n}$.

\begin{figure}[htbp]
\centering
\begin{tabular}{ccc}
\includegraphics[scale=0.2]{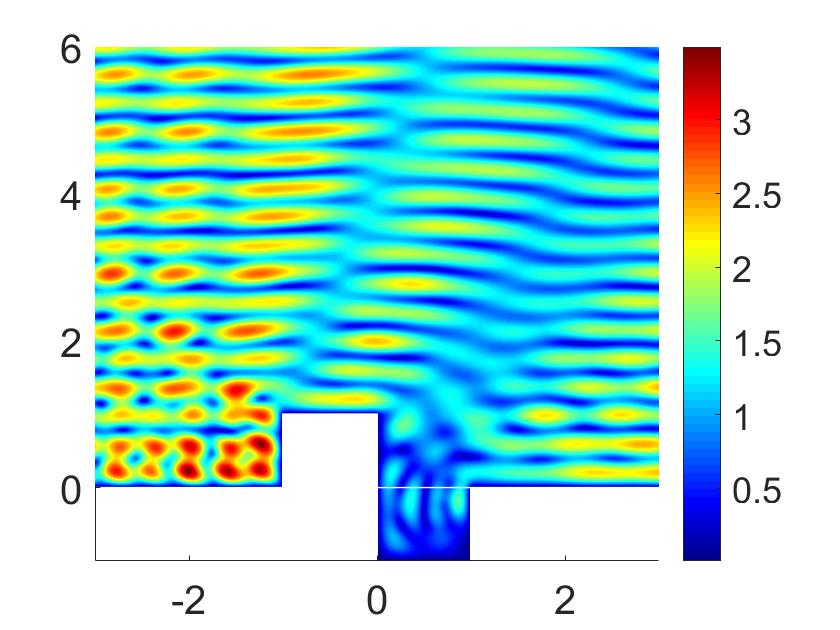} &
\includegraphics[scale=0.2]{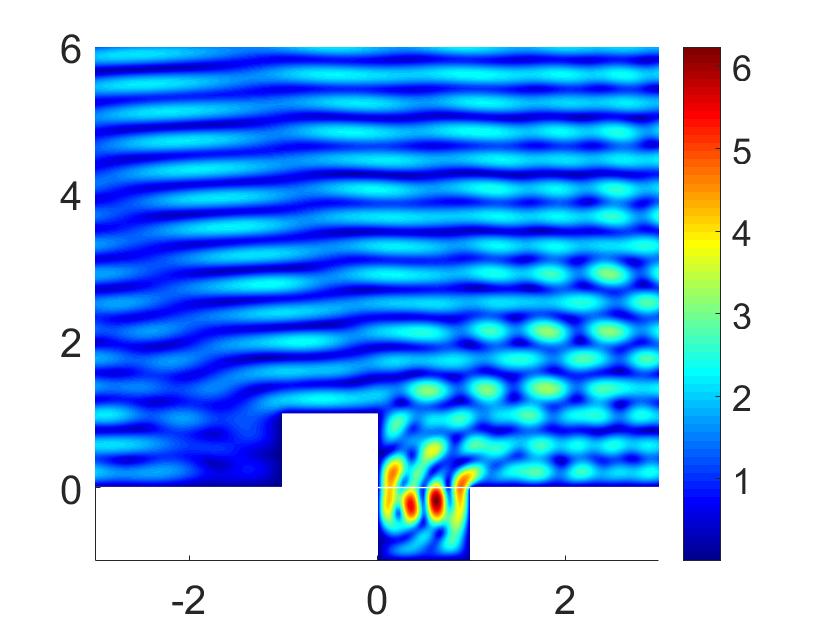} \\
(a) $\theta^{inc}=\frac{\pi}{4}$ & (b) $\theta^{inc}=-\frac{\pi}{4}$ \\
\end{tabular}
\caption{Absolute values of the total field  resulting from the WGF method for the Dirichlet problem of scattering by a locally perturbed surface where $\omega=4\pi$, $A=1+16\lambda_s$; $\epsilon_\infty = $~1E-5.}
\label{Figure2D.4}
\end{figure}

\subsection{2D examples}
In our first example we consider problems of elastic scattering by the
two-dimensional locally-rough surfaces depicted in
Figure~\ref{Figure2D.1}. These include problems of scattering of
bounded scatterers under both Dirichlet and Neumann boundary
conditions over a half-plane (disc-shaped and kite-shaped see
Figure~\ref{Figure2D.1}(a,b)), as well as the local corrugation
depicted in Figure~\ref{Figure2D.1}(c).  In all three cases the
impenetrable (Dirichlet or Neumann) infinite boundary is shown as a
thin black line. Tables~\ref{Table2D.1} and \ref{Table2D.2} display
the relative errors in the total field that result from use of the
proposed WGF method for the Dirichlet and Neumann problems,
respectively, clearly demonstrating the uniform fast convergence of
the proposed approach over wide angular variations, going from normal
incidence to grazing. Figures~\ref{Figure2D.2} and \ref{Figure2D.3}
The near fields for the problem of scattering by the Dirichlet
disc-shaped obstacle and the Neumann kite-shaped obstacle are
presented in Figures~\ref{Figure2D.2} and~\ref{Figure2D.3}, respectively.

We consider next the problem of scattering by a locally-rough surface
containing multiple corners, see Figure \ref{Figure2D.1}(c). In this
example we assumed $\omega=4\pi$, and we utilized a total of twelve
integration patches over the local perturbation, with refinement
exponent $p=4$ at corners, and with window radius
$A=1+16\lambda_s= 9$. Figure~\ref{Figure2D.4} displays the total
fields for the Dirichlet problem with incident angles
$\theta^{inc}=\pi/4$ and $\theta^{inc}=-\pi/4$, respectively. In both
cases the relative error is smaller than 1E-5.

\subsection{3D examples}

\begin{table}[htbp]
\centering
\begin{tabular}{|c|c|c|c|c|c|c|}
\hline
\multirow{2}{*}{$A/\lambda_s$} & \multicolumn{3}{|c|}{Dirichlet problem}  & \multicolumn{3}{|c|}{Neumann problem}\\
\cline{2-7}
& $\theta^{inc}=0$ & $\theta^{inc}=\frac{\pi}{4}$ & $\theta^{inc}=\frac{63\pi}{128}$
& $\theta^{inc}=0$ & $\theta^{inc}=\frac{\pi}{4}$ & $\theta^{inc}=\frac{63\pi}{128}$  \\
\hline
2 & 1.61E-1 & 7.76E-2 & 8.10E-2 & 5.40E-2 & 4.44E-1 & 5.03E-2 \\
3 & 3.03E-2 & 2.37E-2 & 3.44E-2 & 1.49E-2 & 8.24E-2 & 2.03E-2 \\
4 & 5.15E-3 & 4.03E-3 & 7.60E-3 & 3.66E-3 & 5.28E-3 & 1.36E-2 \\
5 & 1.24E-3 & 9.93E-4 & 1.98E-3 & 1.75E-3 & 1.90E-3 & 4.07E-3 \\
6 & 2.27E-4 & 1.75E-4 & 3.51E-4 & 6.19E-4 & 6.38E-4 & 1.94E-3 \\
\hline
\end{tabular}
\caption{Relative errors $\epsilon_\infty$ in the total field  resulting from the  WGF method for Dirichlet and  Neumann problems of scattering by a spherical  obstacle within a half space. }
\label{Table3D.1}
\end{table}

\begin{table}[htbp]
\centering
\begin{tabular}{|c|c|c|c|c|c|c|}
\hline
$\theta^{inc}$ & $A/\lambda_s$ & $M$ & $N_{\mathrm{DOF}}$  & Time (prec.) & Time (1 iter.)& $N_{iter}$ \\
\hline
 & 2 & 22 & $3\times12672$ & 32.92 s & 5.12 s & 29 \\
0 & 4 & 70 & $3\times40320$ & 2.79 min & 1.12 min & 32 \\
 & 6 & 150 & $3\times86400$ & 8.60 min & 5.05 min & 34 \\
\hline
 & 2 & 22 & $3\times12672$ & 33.23 s & 5.07 s & 40 \\
$\frac{\pi}{4}$ & 4 & 70 & $3\times40320$ & 2.84 min & 1.12 min & 42 \\
 & 6 & 150 & $3\times86400$ & 8.55 min & 5.00 min & 45 \\
\hline
 & 2 & 22 & $3\times12672$ & 33.02 s & 5.11 s & 33 \\
$\frac{63\pi}{128}$ & 4 & 70 & $3\times40320$ & 2.87 min & 1.13 min & 33 \\
 & 6 & 150 & $3\times86400$ & 8.57 min & 5.04 min & 34 \\
\hline
\end{tabular}
  \caption{Computing costs required by the WGF method for the Neumann problem of scattering by a spherical obstacle, with GMRES tolerance equal to 1E-4, and with $N=24$ and $N^\beta=100$.}
\label{Table3D.2}
\end{table}

We consider two 3D inclusion types, namely, a sphere and a kite-shaped
obstacle, in both cases over a half space. The total-field relative
errors presented in Table~\ref{Table3D.1} demonstrate the high
accuracy and fast convergence of the proposed 3D WGF method, which is
observed, once again, uniformly for all incidence
angles. Table~\ref{Table3D.2} displays the corresponding computing
costs required by the solver; for definiteness we only present results
for the Neumann case, but the statistics for the corresponding
Dirichlet case are entirely analogous.

\begin{figure}[htbp]
\centering
\begin{tabular}{cc}
\includegraphics[scale=0.15]{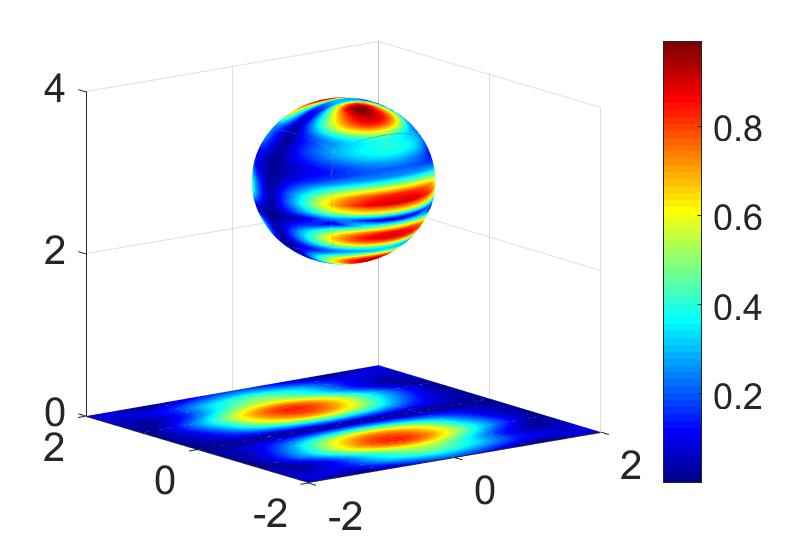} &
\includegraphics[scale=0.2]{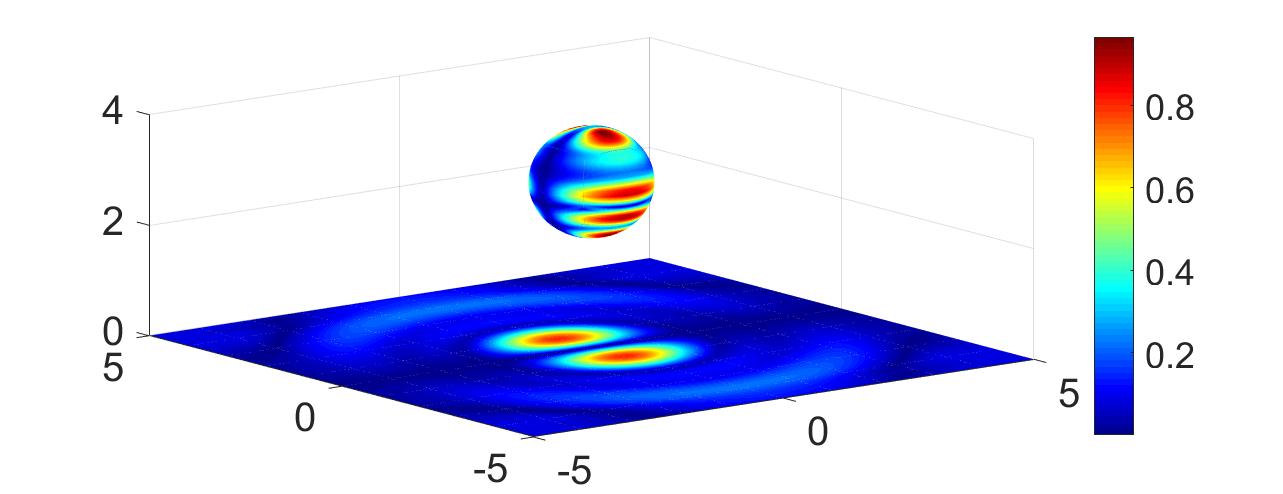} \\
(a) $A=2$, $|u_2|$ & (b) $A=5$, $|u_2|$ \\
\includegraphics[scale=0.15]{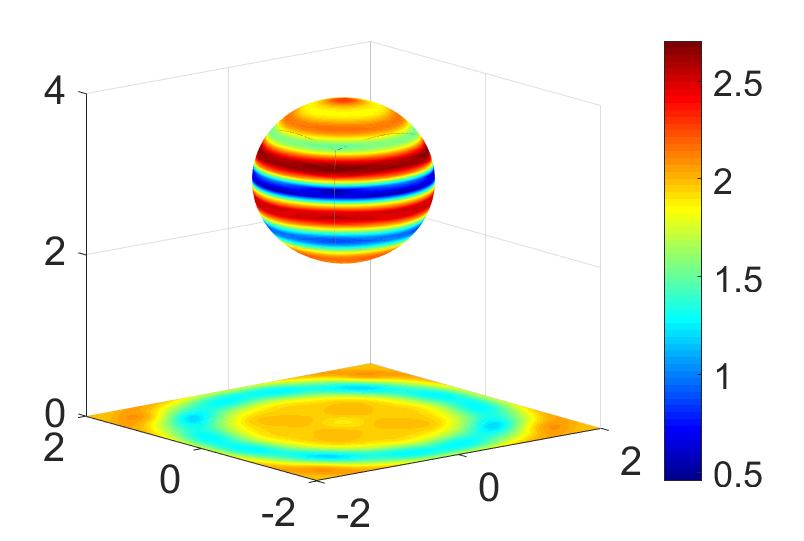} &
\includegraphics[scale=0.2]{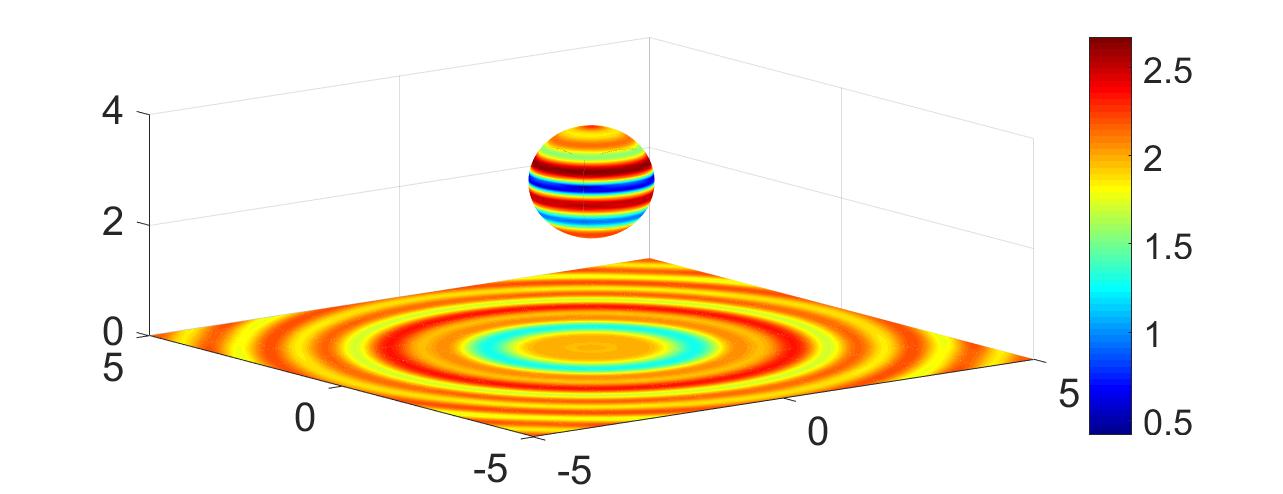} \\
(c) $A=2$, $|u_3|$ & (d) $A=5$, $|u_3|$ \\
\end{tabular}
\caption{Absolute values of the second (a,b) and third (c,d) components of the total field  resulting from the WGF method for the Neumann problem of scattering by a spherical obstacle. The section of the planar interface shown in each case coincides with the windowed region in the plane where the corresponding windowing function $\widetilde{w}_A$ does not vanish. $\theta^{inc}=0$.}
\label{Figure3D.1}
\end{figure}

\begin{figure}[htbp]
\centering
\begin{tabular}{ccc}
\includegraphics[scale=0.14]{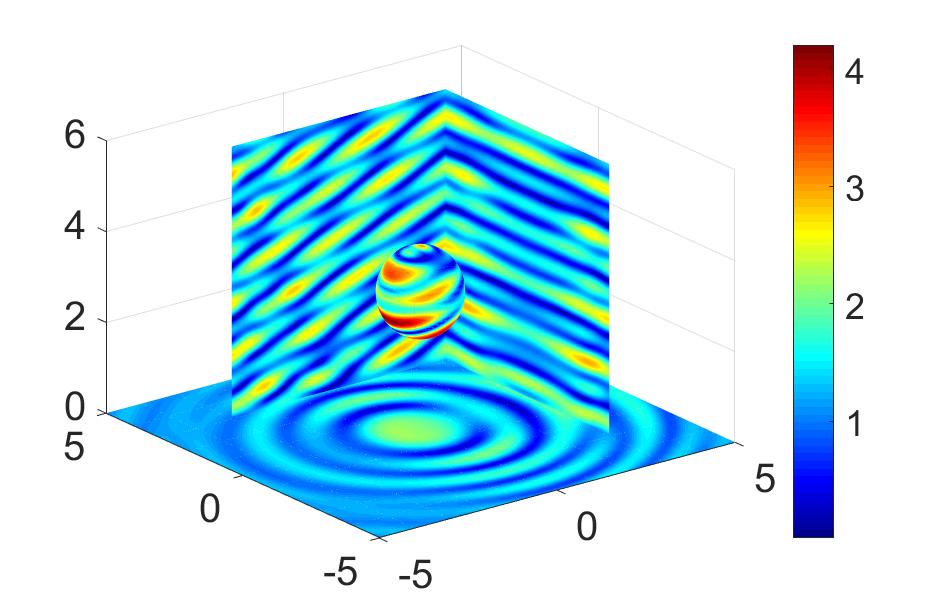} &
\includegraphics[scale=0.14]{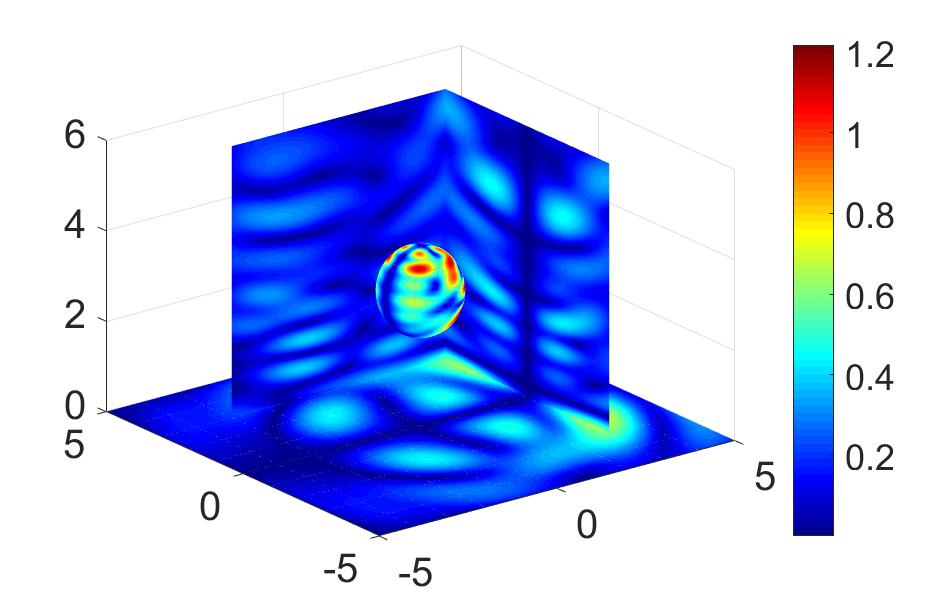} &
\includegraphics[scale=0.14]{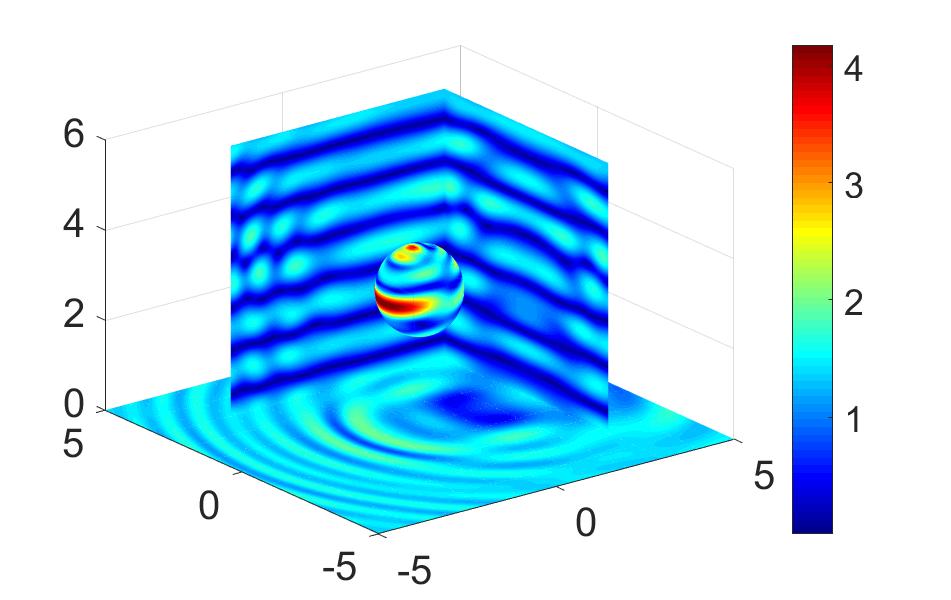} \\
(a) $|u_1|$  & (b) $|u_2|$ & (c) $|u_3|$ \\
\end{tabular}
\caption{Absolute values of the three components of the total field  resulting from the WGF method for the Neumann problem of scattering by a spherical obstacle. The section of the planar interface shown in each case coincides with the windowed region in the plane where the corresponding windowing function $\widetilde{w}_A$ does not vanish. $\theta^{inc}=\pi/4$.}
\label{Figure3D.2}
\end{figure}

Figure~\ref{Figure3D.1} displays the computed values of the total
field for the Neumann problem with $\lambda=\mu=3$ and
$\omega=5\sqrt{3}\pi/2$. These results are consistent with the LGF-based
results presented in~\cite{CB13}, which include a treatment of this
problem but only under $\theta^{inc}=0$ incidence. The LGF evaluation
that is required in the treatment~\cite{CB14}, on the other hand, is
much more expensive, on a per-point basis than the free-space Green
function we use.  A direct truncation of the infinite planar surface
to the square $|x|\le A$  was proposed in~\cite{CBS08,CB13} for
an equation similar to~(\ref{WDBIE1}); as discussed in Sections~\ref{sec:1} and Section~\ref{sec:2} and suggested by the WGF results
in Figure~\ref{Table3.1.1}, however, such approaches lead to significant difficulties as the incidence angles sufficiently depart from normal incidence.

\begin{figure}[htb]
\centering
\begin{tabular}{cc}
\includegraphics[scale=0.2]{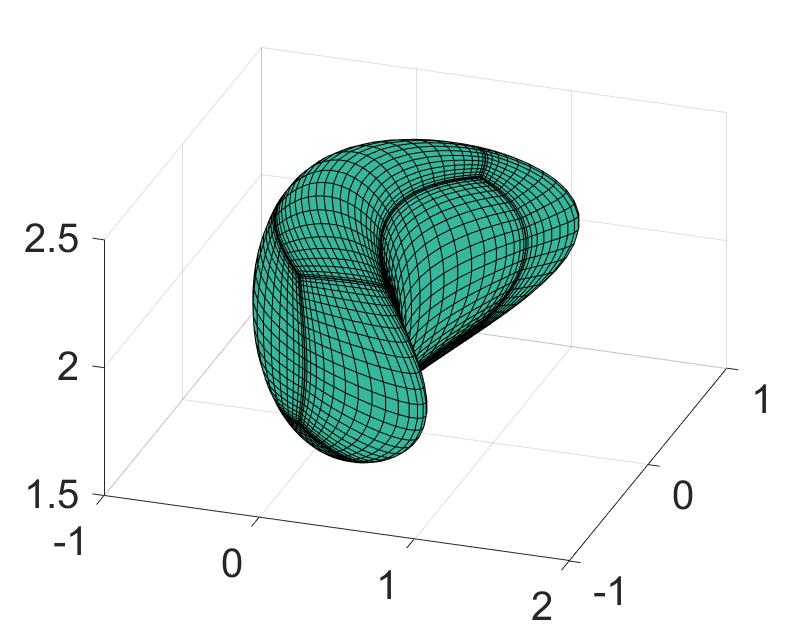} &
\includegraphics[scale=0.2]{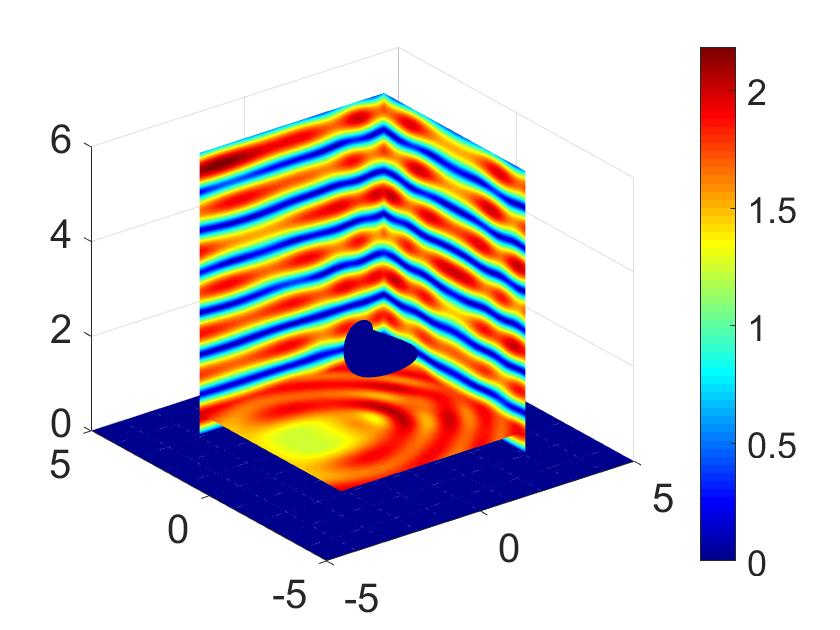} \\
(a) Bean-shaped obstacle & (b) $|u_1|$ \\
\includegraphics[scale=0.2]{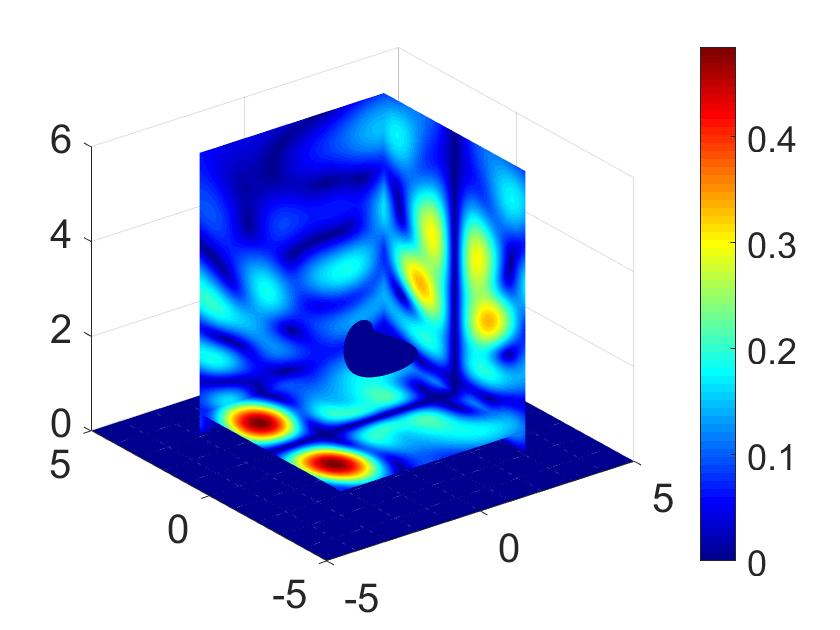} &
\includegraphics[scale=0.2]{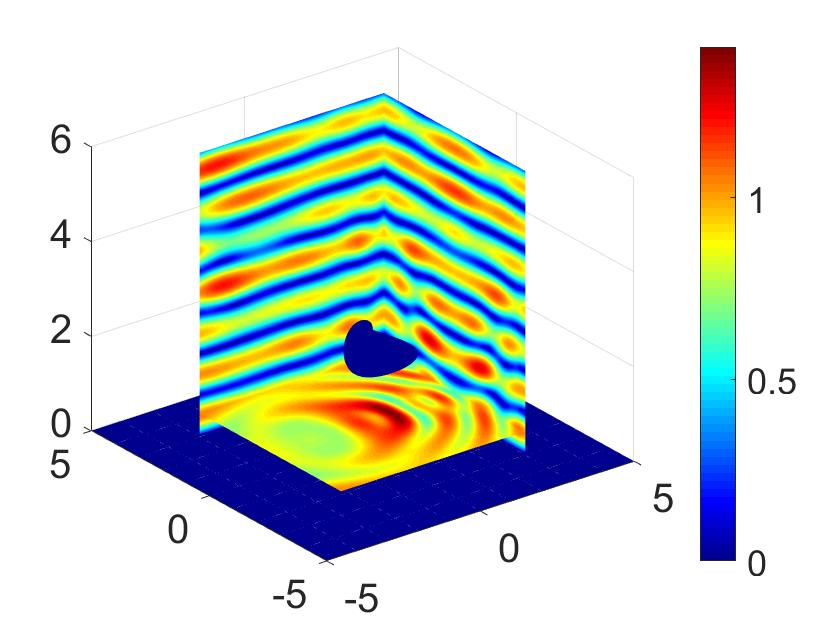} \\
(c) $|u_2|$ & (d) $|u_3|$ \\
\end{tabular}
\caption{Absolute values of the three components of the total field  resulting from the WGF method for the Dirichlet problem of scattering by a Bean-shaped obstacle. The section of the planar interface shown in each case coincides with the windowed region in the plane where the corresponding windowing function $\widetilde{w}_A$ equals to 1. $\theta^{inc}=-\pi/3$.}
\label{Figure3D.3}
\end{figure}

Finally, the total field produced by the WGF method for the Dirichlet
problem of scattering by the bean-shaped obstacle over a half space
displayed in Figure~\ref{Figure3D.3}(a), for a problem with
$\omega=2\pi$, $\theta^{inc}=-\pi/3$ and $A=5\lambda_s$, which was
treated using the algorithmic parameter selections $M=106$, $N=24$ and
$N^\beta=100$, is presented in Figures~\ref{Figure3D.3}(b,c,d). The
relative solution error $\epsilon_\infty$ is smaller than 1E-3 and the
absolute computing time (including precomputation as well as GMRES
iteration and field evaluation) is 1.94h with GMRES tolerance equal to
1E-4. Of course, all of the computing times can be greatly reduced by
means of suitable acceleration method such as those presented
in~\cite{CB14,BK01} and references therein.

\section{Conclusions}

This paper introduced novel WGF methods for the solution of half-space
elastic scattering problems with Dirichlet or Neumann boundary
conditions. Relying on 1)~The free-space Green function, together with
2)~A novel windowed version of the classical elasticity integral
equations, 3)~A novel integral formulation that is uniformly accurate
for all incidence angles, and 4)~Efficient high-order
singular-integration methods, the proposed approach avoids the
expensive evaluation of the elastic layer Green function and, as
demonstrated by a variety of numerical tests, can achieve uniform fast
convergence for all incident angles. Extensions of the WGF approach to
other types of half-space scattering problems, including
e.g. fluid-solid interaction problems with multiple layers~\cite{P97},
Rayleigh wave scattering problems~\cite{AA04}, and scattering problems
with tapered incidence~\cite{T88}, etc., which can be treated by
similar methods, are left for future work.

\section*{Acknowledgments} 
This work was supported by NSF and AFOSR through contracts DMS-1714169
and FA9550-15-1-0043, and by the NSSEFF Vannevar Bush Fellowship under
contract number N00014-16-1-2808.

%\section*{Acknowledgments}
%

\end{document}